\documentclass[10pt,twocolumn,floatfix,aps,pra,superscriptaddress,nofootinbib]{revtex4-2}
\usepackage{amsmath,amssymb,amsthm,mathrsfs,amsfonts,dsfont,amstext} 
\usepackage{textcomp,pbox}
\usepackage[export]{adjustbox}
\usepackage{soul} 
\usepackage{lineno} 
\usepackage{bm,xspace}
\usepackage{dcolumn,booktabs,url}
\usepackage[scaled]{helvet}
\usepackage{sansmath,gensymb}
\usepackage{tikz,graphicx,transparent,color}
\usepackage{multirow}
\usepackage[separate-uncertainty = true]{siunitx}
\usepackage{comment}
\usepackage{physics}
\usepackage{pdfpages}
\renewcommand{\arraystretch}{1.4}

\usepackage[dvipsnames,table,xcdraw]{xcolor}
\usepackage{array}
\usepackage{makecell}
\usepackage{tabularx}
\usepackage{dsfont}
\usepackage{array}
\usepackage{booktabs}

\newcolumntype{L}{>{\raggedright\arraybackslash}X}

\makeatletter
\AtBeginDocument{\let\LS@rot\@undefined}
\makeatother

\newcolumntype{C}[1]{>{\centering\let\newline\\\arraybackslash\hspace{0pt}}m{#1}}
\usepackage{natbib} 

\usepackage[colorlinks=true]{hyperref}
\usepackage{graphicx}

\renewcommand{\arraystretch}{1.0}    

\hypersetup{
     colorlinks   = true,
     citecolor    = red,
     linkcolor    = blue,
     urlcolor     = red     
}

\graphicspath{{./Figures/}}

\newcommand{\SM}{Supplementary Material\xspace}

\makeatletter
\def\maketitle{
\@author@finish
\title@column\titleblock@produce
\suppressfloats[t]}
\makeatother

\usepackage[resetlabels]{multibib}

\newcites{S}{References}

\begin{document}
\newcommand{\TitleName}{Practical blueprint for low-depth photonic quantum computing with quantum dots}

\title{\TitleName}

\author{Ming Lai Chan$^{\dagger}$}
\email{Ming-Lai.Chan@sparrowquantum.com}
\affiliation{Sparrow Quantum, Nordre Fasanvej 215, DK-2000 Frederiksberg, Denmark}
\affiliation{Center for Hybrid Quantum Networks (Hy-Q), The Niels Bohr Institute, University~of~Copenhagen,  DK-2100  Copenhagen~{\O}, Denmark}
\author{Aliki Anna Capatos}
\thanks{These authors contributed equally to this work.}
\affiliation{Quantum Engineering Centre for Doctoral Training, University of Bristol, Bristol, United Kingdom}
\affiliation{NNF Quantum Computing Programme, Niels Bohr Institute, University of Copenhagen, Blegdamsvej 17, DK-2100 Copenhagen Ø, Denmark}
\author{Peter Lodahl}
\affiliation{Sparrow Quantum, Nordre Fasanvej 215, DK-2000 Frederiksberg, Denmark}
\affiliation{Center for Hybrid Quantum Networks (Hy-Q), The Niels Bohr Institute, University~of~Copenhagen,  DK-2100  Copenhagen~{\O}, Denmark}
\author{Anders Søndberg Sørensen}
\affiliation{Center for Hybrid Quantum Networks (Hy-Q), The Niels Bohr Institute, University~of~Copenhagen,  DK-2100  Copenhagen~{\O}, Denmark}
\author{Stefano Paesani}
\email{stefano.paesani@nbi.ku.dk}
\affiliation{Center for Hybrid Quantum Networks (Hy-Q), The Niels Bohr Institute, University~of~Copenhagen,  DK-2100  Copenhagen~{\O}, Denmark}
\affiliation{NNF Quantum Computing Programme, Niels Bohr Institute, University of Copenhagen, Blegdamsvej 17, DK-2100 Copenhagen Ø, Denmark}
\date{\today}

\maketitle
\section*{Abstract}
Fusion-based quantum computing is an attractive model for fault-tolerant computation based on photonics requiring only finite-sized entangled resource states followed by linear-optics operations and photon measurements. Large-scale implementations have so far been limited due to the access only to probabilistic photon sources, vulnerability to photon loss, and the need for massive multiplexing. Deterministic photon sources offer an alternative and resource-efficient route. By synergistically integrating deterministic photon emission, adaptive repeat-until-success fusions, and an optimised architectural design, we propose a complete blueprint for a photonic quantum computer using quantum dots and linear optics. It features time-bin qubit encoding, reconfigurable entangled-photon sources, and a fusion-based architecture with low optical connectivity, significantly reducing the required optical depth per photon and resource overheads. We present in detail the hardware required for resource-state generation and fusion networking, experimental pulse sequences, and exact resource estimates for preparing a logical qubit. We estimate that one logical clock cycle of error correction can be executed within microseconds, which scales linearly with the code distance. We also simulate error thresholds for fault-tolerance considering a full catalogue of intrinsic error sources found in real-world quantum dot devices. Our work establishes a practical blueprint for a low-optical-depth, emitter-based fault-tolerant photonic quantum computer.


\section*{Introduction}
Building a useful quantum computer rests on designing a practical framework that integrates platform constraints, error correction, and scalable control systems. A blueprint for such a framework must bridge theory and hardware capabilities and describe precise physical component requirements, their integration and overhead, and performance targets under realistic conditions that may guide experimental progress. 

Currently, blueprints tailored to different physical platforms~\cite{Bluvstein2026, xanadu_blueprint, percolation_arch, gutiérrez2025, Larsen2021, Auger2017, Kim2025faulttolerant, Pino2021, cat_codes_architecture} have been proposed, addressing a range of aspects including specific hardware design, physical error modelling and resource estimation.

Among the candidate platforms, photonic quantum computing offers fast gate times, scalability, modularity, and room-temperature operation, but faces three stumbling blocks. First, resource-state generation based on spontaneous parametric down conversion is highly inefficient due to the low success probability~\cite{sahay_tailoring_2022}. Since each physical qubit in a resource state is usually encoded in multiple photons, which requires additional probabilistic gates, the overall success rate is very low. Preparing a logical qubit of code distance $d=20$ is estimated to require $2d^3=16,000$ such states~\cite{bartolucci_fusion-based_2023}. Alternatively, continuous-variable systems use bright beams that enable deterministic entanglement generation~\cite{Lund2008}, yet these schemes are highly sensitive to loss and require non-Gaussian photon operations that are typically implemented probabilistically by photon subtraction. Second, current photonic architectures typically rely heavily on complex routing involving fibre delays and multiplexing. This introduces large optical depths in resource-state generation and/or photon measurement circuits, resulting in many loss channels which hinder scalability~\cite{bartolucci_fusion-based_2023, xanadu_blueprint, bartolucci2021creationentangledphotonicstates, bartolucci2021switchnetworksphotonicfusionbased}. Third, the current state of photonic hardware~\cite{Alexander2024} places photon loss rates well above the thresholds required for fault-tolerant fusion-based quantum computation (FBQC)~\cite{bartolucci_fusion-based_2023}, and while existing methods can tolerate moderate photon loss (around 20\%~\cite{Omkar2022, Bell_2023_optimizing, Song2024, pankovich_high_2023, Lobl2024losstolerant, Thomas2024, Borregaard2020, economou2017, deGliniasty2024spinopticalquantum, Auger2018}), strategies for improving this tolerance without enormous overhead in the photon number per resource state~\cite{psiq_loss_comparison, litinski2025blocklets} have not been identified. 

\begin{figure*}[hbtp!]
	\includegraphics[width=0.82\linewidth, trim=0.cm 0cm 0.cm 0.0cm,clip]{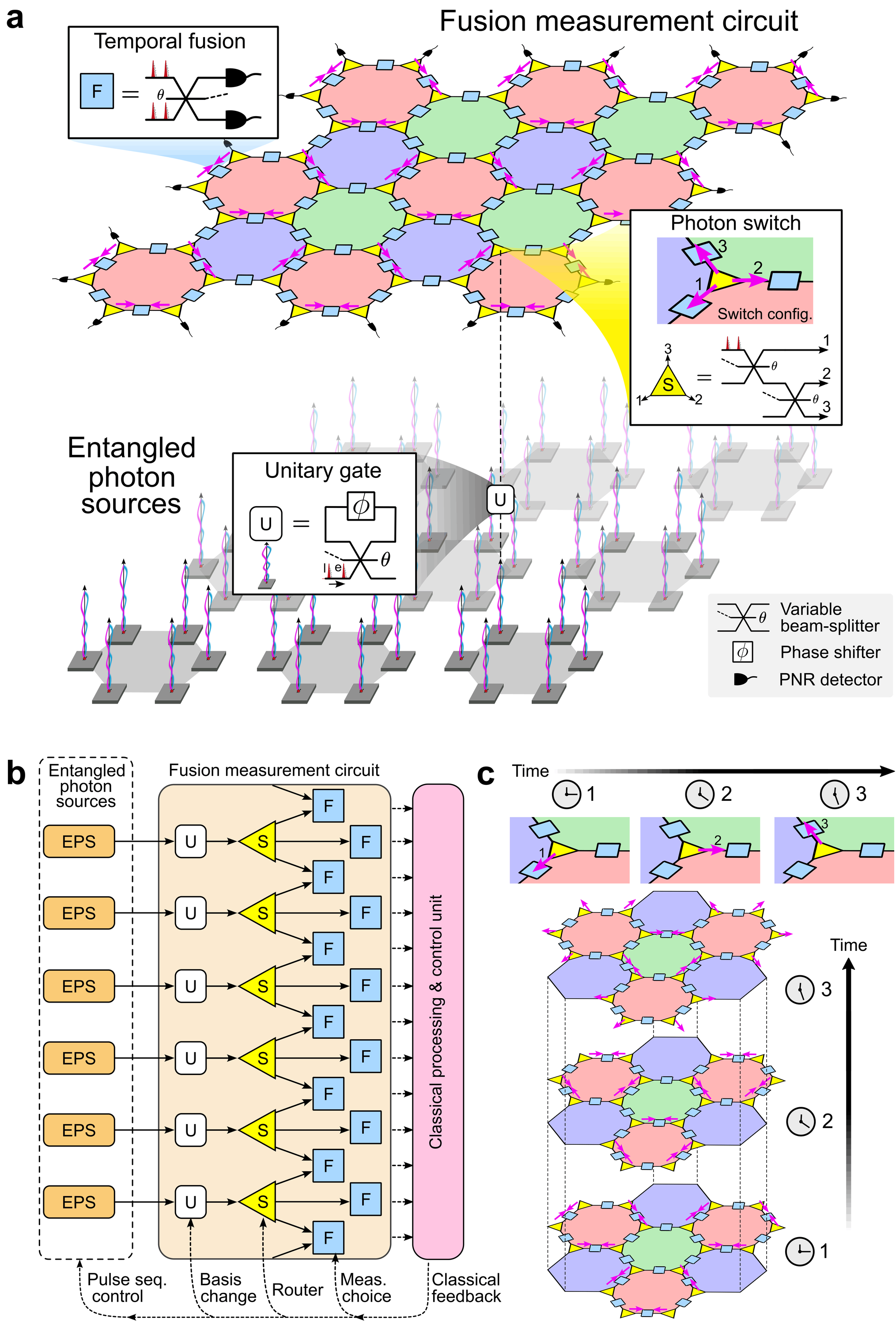}
    \centering
    \caption{\textbf{Emitter-based photonic quantum computing blueprint.} \textbf{(a)} An array of deterministic entangled-photon source (EPS) units sends photonic resource states (dashed line) to a planar chip with a fusion measurement circuit to grow and measure the foliated Floquet colour code (FFCC) cluster state of code distance $L=3$. Each EPS unit sends a time-bin encoded photon through a unitary gate (white square) for state projection, a photon switch (yellow triangle) to route it to one of three optical paths, and finally, a fusion gate (blue square). The lattice is three-colourable, where fusion measurements within cells of the same colour form stabiliser checks of the colour code.
    \textbf{(b)} Blueprint overview depicting the relationship between various components. Fusion outcomes are processed by a classical control unit, which depending on the outcome, modifies the EPS units and active optical components. \textbf{(c)} Photon switch order with time to perform fusions on different lattice layers.}
    \label{fig:overview}
\end{figure*}

To overcome these obstacles, recent photonic architectures leverage deterministic photon emitters to reliably produce high-fidelity entangled photonic resource states~\cite{paesani_high-threshold_2022, deGliniasty2024spinopticalquantum, Pettersson2025, Wein2024}, and adaptive repeat-until-success (RUS) fusion gates to significantly boost tolerance to photon loss~\cite{hilaire2024enhancedfaulttolerancephotonicquantum, Chan2025}. Foundational experiments have already successfully demonstrated both on-demand resource-state generation~\cite{Thomas2022, Meng2024, Huet2025, Cogan2023} and fusion gates~\cite{Thomas2024, meng2023photonicfusionentangledresource}, the essential building blocks for FBQC. 

With these theoretical and experimental advances, a critical next step is to integrate these techniques into a practical blueprint for emitter-based photonic architectures. However, several key aspects must be considered. First, current architectures primarily assume path-encoded photons, which may not align with the photonic degree of freedom in emitter-generated resource states. This choice of photon encoding influences both circuit layout and overhead. Second, practical architectures must account for the specific error mechanisms associated with different emitter platforms. Third, realising adaptive fusion operations requires detailed analysis of hardware constraints, including speed and timing requirements, control sequences, and pulse synchronisation across electronic and photonic hardware. Finally, a complete assessment should quantify the computation time and physical overhead of the full architecture, not merely resource-state generation~\cite{Wein2024}.

Here, we present a comprehensive, practical blueprint for an emitter-based photonic quantum computing architecture that directly addresses these challenges. Building upon previous research on deterministic entangled-photon sources (EPS)~\cite{Meng2024}, linear-optics temporal fusions~\cite{meng2023photonicfusionentangledresource}, and the synchronous foliated Floquet colour code (sFFCC) fusion network~\cite{Chan2025}, our proposal unifies these three elements in an architectural design with several unique features. By pairing temporal fusions with the natural time-bin photon encoding of emitter-generated resource states, we simplify the architecture substantially. Specifically, the connection between an EPS unit and the fusion measurement circuit reduces to a single optical path, minimizing optical depth and loss channels. In the sFFCC framework, photons emitted from an array of EPS units pass through unitary gates, photon switches, and fusion gates (Fig.~\ref{fig:overview}a-b). The emitted photons are synchronously fused, thereby eliminating the need for fibre delays between each EPS unit from resource-state generation to detection. The sFFCC fusion network performs fusions layer by layer to generate a cluster state protected by the foliated Floquet colour code~\cite{Kesselring2024,paesani_high-threshold_2022}. As the cluster state is constructed, each resource state interacts with only three neighbouring states under a fixed routing configuration (Fig.~\ref{fig:overview}c). Thus, each photon traverses a minimal number of optical components --- only five active phase shifters and up to eight passive beamsplitters --- making this minimal-depth architecture an essential asset that brings real-world implementations within reach. 

The architecture utilises adaptive fusions that employ feedforward control directly on excitation pulses of EPS units, a method unique to emitters with reconfigurable pulse sequences, which could render the fusion gates \textit{fully passive}. We then provide a thorough description of the necessary hardware and timing constraints.  

Furthermore, we quantify all major resource costs of the full architecture. This includes the number of emitters, detectors, active and passive optical components, and the time to complete one logical quantum error correction (QEC) clock cycle. 

In addition, while previous works~\cite{deGliniasty2024spinopticalquantum, hilaire2024enhancedfaulttolerancephotonicquantum, Chan2025} considered hardware-agnostic noise models, we perform a full analysis of experimental noise sources native to the semiconductor quantum dot (QD) platform, show how these complex error mechanisms can be effectively captured by simple error models, and simulate corresponding error correction thresholds. 

Our work outlines a clear and feasible path toward large-scale photonic quantum computing with quantum emitters, presenting a realistic architectural blueprint specified at the hardware level, which we hope will inspire and guide future experimental implementations.


\section*{Results}
\subsection{Fusion-based quantum computing architecture}
\label{fbqc}
Well-suited to photonic systems, FBQC~\cite{bartolucci_fusion-based_2023} is a measurement-based quantum computing model~\cite{Raussendorf2001} comprised of \textit{fusing} smaller photonic resource states together via two-qubit entangling measurements, to effectively grow and measure a cluster state in a single step. Quantum information is processed by performing two-qubit destructive measurements, called fusion gates, on these small resource states, which allows syndrome data to be extracted for error correction, where the algorithm is implemented by controlling the measurement settings of fusion gates. We begin this section by reviewing relevant concepts of FBQC, including the physical fusion between two photons, encoded fusion, and fusion networks.

A physical type-II photonic fusion gate~\cite{Browne2005,bartolucci_fusion-based_2023} is a destructive two-qubit measurement, which upon success, projects two input photons onto the Bell-state basis to generate entanglement. A fusion gate is probabilistic and succeeds with $50\%$ probability in the absence of errors, since entanglement is heralded by conditioning on certain click patterns with photon-number-resolving (PNR) detectors. 

\begin{figure}[hb]
	\includegraphics[width=\linewidth, trim=0.cm 0cm 0.cm 0.0cm,clip]{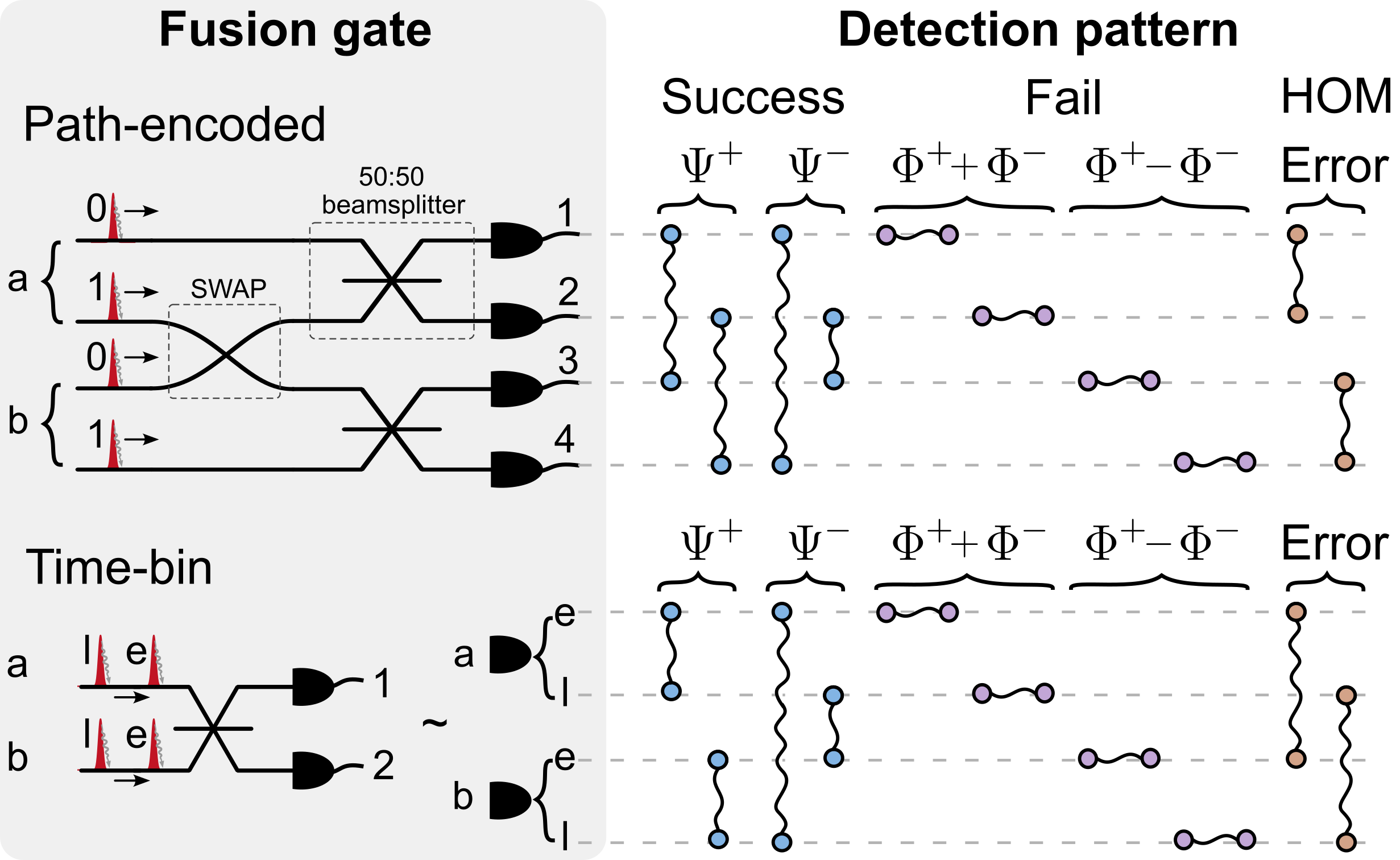}
    \centering
    \caption{\textbf{Linear-optic fusion gates for path and time-bin encodings.} Each fusion gate has input qubits $a$ and $b$. Circles connected by a wiggly line represent a two-photon coincidence click recorded by the detector(s). Different detection click patterns indicate that the fusion has resulted in either a success (Bell-state projection into $\Psi^\pm$), or failure (computational bases), or failure but with an error due to imperfect Hong-Ou-Mandel interference. In the case of photon loss, only one or no detector click is recorded.}
    \label{fig:type}
\end{figure}

Fig.~\ref{fig:type} shows two types of physical $\{XX,ZZ\}$ fusion for photons encoded in different degrees of freedom. In path-encoded (spatial) fusion, two photonic qubits ($a$ and $b$) are sent through four optical paths in a fusion gate. A photon in the first (third) path represents the logical state $\ket{0}$ and a photon in the second (fourth) represents $\ket{1}$. The fusion gate consists of linear-optic elements including a SWAP gate, a passive beamsplitter, and four photodetectors. 
Similarly, photonic fusion can be performed with the time-bin encoding~\cite{meng2023photonicfusionentangledresource}. It mirrors spatial fusion but operates in the time domain, using only two optical paths, where the $\ket{0}$ and $\ket{1}$ states of a photonic qubit propagate along the same path. The logical information is encoded in the photon's arrival time, with $\ket{e}\equiv\ket{0}$ and $\ket{l}\equiv\ket{1}$. Temporal fusion is especially attractive for FBQC due to its reduced overhead; notably, it requires only half the number of photodetectors. 

Below, we summarise the four outcomes of a fusion:
\begin{itemize}
    \item Fusion success. A successful fusion between photonic qubits $a$ and $b$ measures the commuting operators $X_a X_b$ and $Z_a Z_b$, where $X,Z$ are Pauli operators, projecting them onto a Bell state. This is referred to as a $\{XX,ZZ\}$ fusion.
    \item Fusion failure. Only one detector lights up but it resolves two clicks in the same mode due to photon bunching. For a $\{XX,ZZ\}$ fusion, this effectively performs single-qubit measurements $Z_a I_b$, $I_a Z_b$. The $Z_a Z_b$ fusion outcome can still be recovered by multiplying these measurement outcomes, thus it is referred to as the \emph{fusion failure basis}~\cite{bartolucci_fusion-based_2023}. In this work, $\{ZZ,XX\}$ fusion is instead used, where a Hadamard rotation is applied to each photon before the fusion circuit in Fig.~\ref{fig:type}, to retrieve $XX$ upon failure.
    \item Fusion error. When the input photons are not indistinguishable, they do not exhibit perfect Hong-Ou-Mandel photon interference at the beamsplitter, resulting in a coincidence click between two undesired modes. These outcomes are interpreted as a fusion failure, since $Z_a Z_b$ can be recovered.
    \item Fusion erasure. If one or both photons involved in a fusion are lost, either one click in the same mode or no click is registered. As such, both $X_a X_b$ and $Z_a Z_b$ outcomes are erased.
\end{itemize} 

Encoded fusion refers to a fusion between two encoded qubits. The objective is to mitigate the failure of fusing two resource states and to protect against photon loss by introducing redundancy in the fusions. Specifically, if one of the input photons to a physical fusion is lost, both measurement outcomes $XX$ and $ZZ$ cannot be recovered since this requires simultaneous measurement of both photons. With an $N$-qubit repetition code, the measurement operators of an encoded fusion can be written as~\cite{Chan2025}

\begin{align}
    \overline{X}_{a} \overline{X}_{b}=\bigotimes^{N}_{j,k=1}X_{a_j}X_{b_k};\quad\overline{Z}_a \overline{Z}_b=Z_{a_j}Z_{b_k},
\end{align}
where $a_j$ ($b_k)$ is the index of $j$-th ($k$-th) photon of encoded qubit $a$ ($b$). These operators can be obtained by attempting $N$ physical fusions. As such, if one of the photons in the first physical fusion is lost, it is still possible to recover $\overline{ZZ}=Z_{a_j}Z_{b_k}$ provided that there is at least one fusion success between any pair of photons $a_j$ and $b_k$ within $N$ attempts without loss. To recover $\overline{XX}$, however, all $2N$ photons must survive. Encoded fusion is particularly powerful in suppressing fusion failure; since only one $ZZ$ outcome is needed to obtain $\overline{ZZ}$, the probability of obtaining $\overline{ZZ}$ approaches unity for large $N$, assuming no photon loss. When the failure basis is set to $XX$, even if all physical fusions fail, the $\overline{XX}$ outcome may still be recovered. 

So far, we have described a type of encoded fusion that utilises a static repetition code in which the code size $N$ is constant, thus each encoded fusion requires $2N$ physical qubits. This is sub-optimal, as it is likely that one or both encoded measurement operators can be recovered within the first few physical fusion attempts. Continuing the remaining attempts is redundant and reduces the chance to recover $\overline{XX}$. A more resource-efficient and loss-tolerant type of encoded fusion is the repeat-until-success (RUS) fusion~\cite{Chan2025,deGliniasty2024spinopticalquantum}, which varies $N$ depending on the previous physical fusions. 

A RUS encoded fusion is completed when either one or both encoded measurement operators are recovered, or a maximum number of attempts is reached. In particular, when the first physical fusion already succeeds, the encoded fusion terminates as both measurement operators $\overline{XX}=XX$, $\overline{ZZ}=ZZ$ are obtained. An important strategy that can be exploited with RUS fusions is that, when a physical fusion suffers from photon loss, the next attempt can be ``biased" in $ZZ$ by changing the fusion failure basis, so single-qubit measurements are performed instead of fusion~\cite{Chan2025}. This deterministically recovers $\overline{ZZ}=ZZ$, suppressing the probability that both $\overline{XX}$ and $\overline{ZZ}$ are erased. To implement RUS fusion, an adaptive feedback needs to be established from the classical control unit (which interprets the fusion measurement outcome) to the EPS unit and fusion measurement circuit. This will be covered in detail in Sec.~\ref{sec:classical}. 

An FBQC architecture, alternatively called a \textit{fusion network}, describes the arrangement of fusion measurements required to construct a logical qubit and perform quantum computation. The architecture adopts certain periodic structures, like the Raussendorf-Harrington-Goyal (RHG) lattice~\cite{bartolucci_fusion-based_2023}, forming an error correction code that improves resilience of the logical qubit to error and loss. Fusions are performed between smaller \emph{cluster states} of entangled photons, called resource states, which then builds the full lattice.

\begin{figure}[hbtp!]
    \centering
	\includegraphics[width=0.95\linewidth, trim=0.cm 0cm 0.cm 0.0cm,clip]{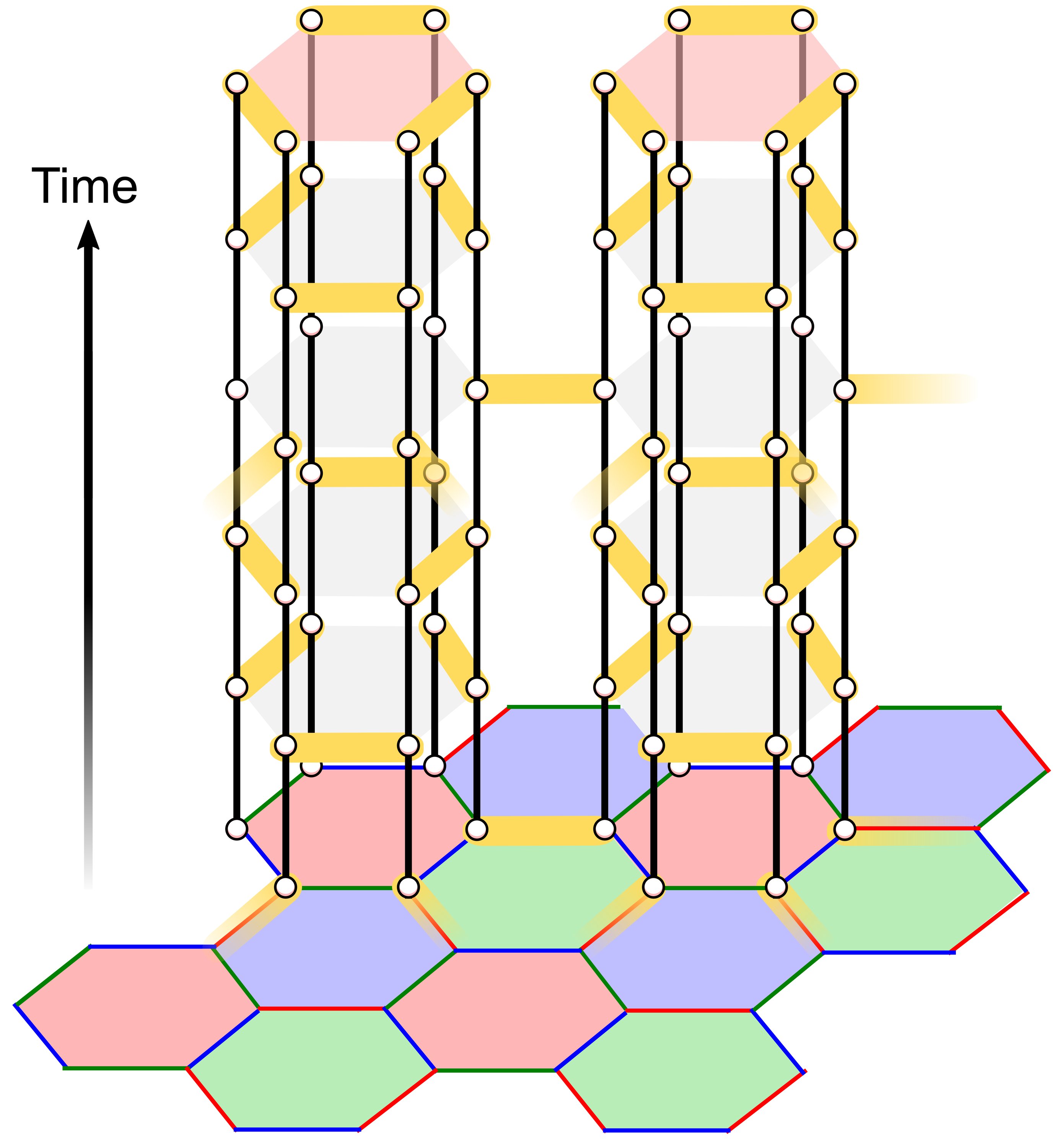}
    \caption{\textbf{Synchronous FFCC fusion network.} The foliated Floquet colour code (FFCC) lattice is decomposed into fusions (orange links) between linear cluster states (white circles connected by black solid lines) at the vertices. Horizontal entanglement links of the lattice are generated by performing synchronous fusions layer by layer. Vertical entanglements in time (black lines) are provided by entangled photons (white circles), emitted sequentially. Information therefore propagates in \textit{time} along the lattice.}
    \label{fig:fusion_network}
\end{figure}

The lattice we consider is the foliated Floquet colour code (FFCC) lattice (Fig.~\ref{fig:fusion_network}), constructed by foliating a two-dimensional Calderbank–Shor–Steane (CSS) code into a three-dimensional cluster state in the measurement-based picture~\cite{paesani_high-threshold_2022}. 

Recent works~\cite{paesani_high-threshold_2022,Chan2025} demonstrate that the FFCC lattice can be viewed as a network of one-dimensional cluster-state chains. In this construction, each data qubit occupies a vertex of a three-colourable hexagonal lattice (Fig.~\ref{fig:fusion_network}), with edges representing successful fusion measurements that connect qubits. Each such edge adopts the colour of the two faces it separates, where no two adjacent faces share the same colour. We then “Floquet‑ify” the structure by cycling through time‑like layers, each time measuring all edges of a single colour. This dynamic schedule encodes the logical information across time. Measuring the edges in one layer performs a check operator, and combining checks across successive layers of the same colour forms a stabiliser operator. Without fusion errors, these stabilisers yield the \(+1\) eigenvalue; a \(-1\) outcome, the \emph{syndrome}, indicates an error in that cell. Together, these syndromes produce a syndrome graph, which is used to detect and track errors on the logical qubit.

Besides the natural congruency of two-qubit fusion measurements with the two-body check measurements of the code, a remarkable feature of sFFCC is its compatibility with deterministic photon emitters that generate entangled photons sequentially in time. This enables the creation of vertical entanglement links, i.e., linear cluster states, within the FFCC lattice (Fig.~\ref{fig:fusion_network}). Horizontal entanglement is thus established by performing fusions between photons from neighbouring linear cluster states. Since each fusion requires two photons to arrive simultaneously at a beamsplitter, any two neighbouring emitters can be synchronised to emit photons concurrently into the fusion gate. This synchrony allows for RUS fusions without the need for fibre delays between EPS units.

We note that the sFFCC architecture~\cite{Chan2025} and the $\text{M}_{\text{ZZ}}$-SPOQC architecture introduced in Ref.~\cite{dessertaine2026}
are mathematically equivalent. At the physical level, they both build upon entangled photon sources from quantum emitters, RUS operations, and Floquet codes with emitters located at the lattice vertices.
The distinction lies in perspective: Ref.~\cite{Chan2025} presents a purely photonic, fusion-based picture in which the spin degrees of freedom are traced out after photon emission and noise is mapped directly onto measurement errors in the fusion outcomes, whereas the emitter-centric picture in Ref.~\cite{dessertaine2026} describes the computation as photon-mediated gates between emitter spins and performs circuit-level simulations using Stim~\cite{stim}. In this work, we retain the FBQC picture in~\cite{Chan2025} for consistency, as it provides the most natural framework for our hardware and error analysis.

\subsection{Blueprint hardware}
\label{architecture}
In this section, we describe the hardware components required to implement a logical qubit system for fusion-based quantum computing, which consists of on‐demand resource-state generation, on-chip fusion measurements, and classical processing.

\begin{figure*}[hbtp!]
	\includegraphics[width=\linewidth, trim=0.cm 0cm 0.cm 0.0cm,clip]{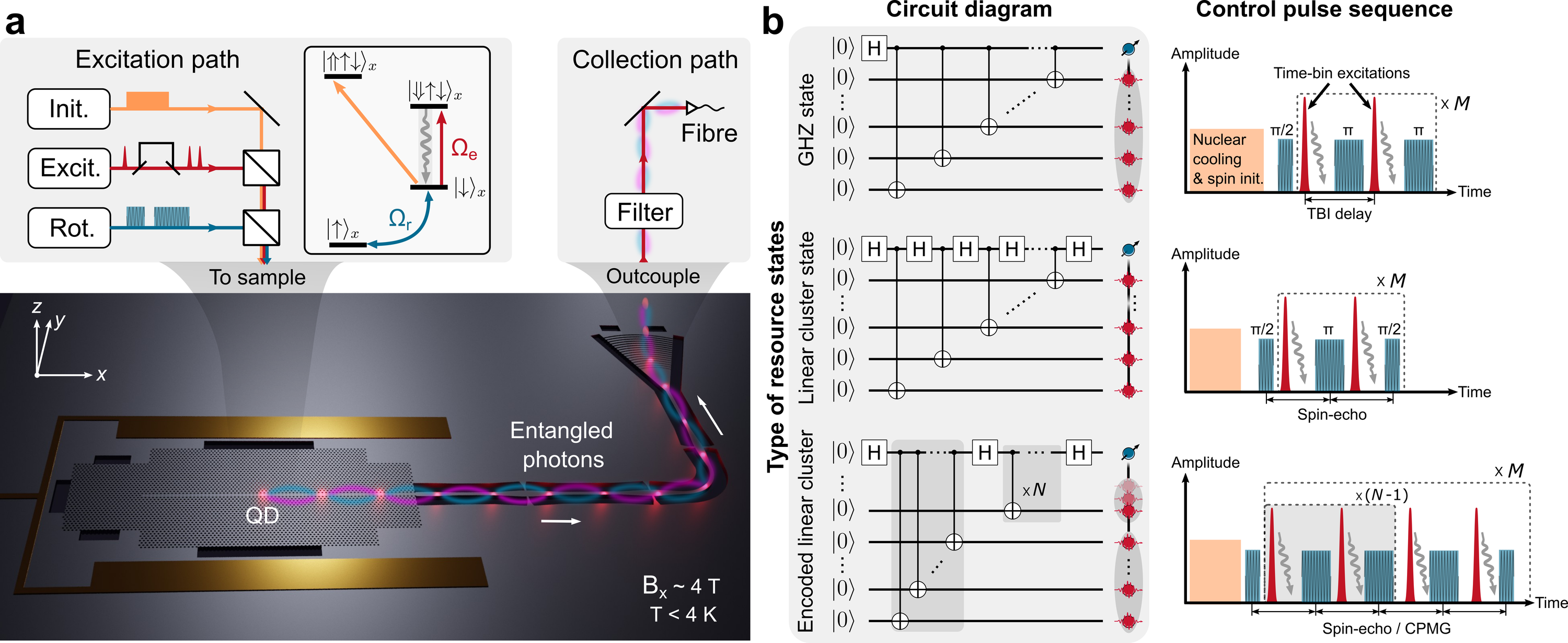}
    \centering
    \caption{\textbf{Entangled-photon source (EPS).} \textbf{(a)} Schematic of an EPS unit based on quantum dots (QD) embedded in photonic-crystal waveguides. Optical pulses for spin initialisation, time-bin excitation (Rabi frequency $\Omega_e$) and spin rotation ($\Omega_r$) are directed toward an EPS chip located inside a cryostat, which supplies a high magnetic field. The top centre panel shows the required QD energy level diagram, with (shaded) cycling transition $\ket{\Downarrow\uparrow\downarrow}_x\to\ket{\downarrow}_x$. The EPS unit emits photons which are outcoupled through an outcoupling grating and into an optical fibre. \textbf{(b)} Quantum circuits and time-bin protocols for generating photonic resource states. A time-bin photon is created by two optical excitation pulses (separated by a time delay) interleaved with a spin $\pi$-rotation pulse. As shown in the top (middle) panel, a GHZ (linear cluster) state is generated by applying a spin $\pi$ ($\pi/2$) pulse after creating a time-bin photon. The resource state considered is the encoded linear cluster state (bottom panel). It is a linear cluster state with $M$ encoded qubits, where each encoded qubit consists of $N$ photons resembling a GHZ state. A spin-echo or Carr-Purcell-Meiboom-Gill (CPMG) sequence is inherent to the generation protocol.}
    \label{fig:rsg}
\end{figure*}

The objective of a resource-state generator (RSG) is to produce entangled-photon resource states~\cite{bartolucci_fusion-based_2023}. An entangled-photon source (EPS) can generate high-quality resource states on demand per logical clock cycle. In general, an EPS unit operates by applying a sequence of optical pulses to a quantum emitter embedded in a nanostructure, enabling strong light–matter interaction and efficient photon collection. 
In this section, we use the semiconductor quantum dot (QD) platform as an example to describe physical components of the EPS unit, its operational principles, and protocols for generating resource states.

The physical platform we consider for the EPS uses a negatively charged QD embedded in a single-sided photonic-crystal waveguide to emit strings of photons entangled with its spin states upon periodic pulsed excitations (Fig.~\ref{fig:rsg}a). In the following, we summarise some of the detailed considerations of actual experimental implementations, including references for further details. As quantum emitters, we consider molecular-beam-epitaxy grown QDs in III-V semiconductor materials (notably InAs in GaAs and GaAs in AlGaAs) that have proven to be the highest quality photon sources available~\cite{Uppu2021}. The QD heterostructure constitutes a p-i-n diode enabling charging of the QD state with electrons or holes, charge noise suppression, and electrical tuning of QD resonance~\cite{lodahl2015}. At the bottom of the diode, a series of alternating AlAs and GaAs layers is grown to build a distributed Bragg reflector to increase photon collection efficiency. 

Under a strong $x$-plane Voigt magnetic field ($B_x\approx 4$~T), the charged QD exhibits a four-level energy system, accessing four optical transitions and two Zeeman-split electronic ground states $\ket{\uparrow}_x$ and $\ket{\downarrow}_x$ (Fig.~\ref{fig:rsg}a). Upon optical excitation, an electron-hole pair is created, which forms two excited states ($\ket{\Uparrow\uparrow\downarrow}_x$ and $\ket{\Downarrow\uparrow\downarrow}_x$), called trions~\cite{lodahl2015}. A single-sided photonic-crystal waveguide is designed to modify the local density of optical states, in which photon emission is greatly enhanced along a preferential direction of the waveguide, resulting in near-unity single-photon coupling efficiency~\cite{Arcari2014} and highly cycling optical transitions~\cite{Appel2021} ($\ket{\Uparrow\uparrow\downarrow}_x\to \ket{\uparrow}_x$ and $\ket{\Downarrow\uparrow\downarrow}_x\to \ket{\downarrow}_x$). The photonic chip is operated inside a cryostat under $T<4$~K to suppress phonon-induced dephasing, with superconducting magnets supplying the required external in-plane magnetic field. The QD is optically excited from the top of the chip using a wide field-of-view confocal microscope, wherein optical pulses for spin initialisation, time-bin excitation, and spin rotation are applied. Entangled photons emitted from the QD can then be collected using a specially engineered grating outcoupler~\cite{Zhou2018} through the same microscope objective, filtered to remove residual phonon sidebands~\cite{lodahl2015}, and imaged onto a single-mode optical fibre.

We now elaborate on the essential conditions for optimal operation of an EPS unit:
\begin{itemize}
    \item Optical excitation pulses should be shorter than the emitter lifetime (to avoid re-excitation) but spectrally narrower than the splitting between the two (shaded) cycling transitions.
    A wider spectral width leads to unwanted driving of another cycling transition. The effect of sub-optimal optical excitations is discussed in Sec.~\ref{sec:errormodels}. 
    Ideally, one of the cycling transitions (e.g., $\ket{\downarrow}_x\to\ket{\Downarrow\uparrow\downarrow}_x$) is driven with an optical $\pi$-pulse of Rabi frequency $\Omega_e$. 
    \item The spin transition should be driven with a fast Rabi frequency $\Omega_r$ to avoid driving Hartmann-Hahn resonances and ensure a high spin rotation Q-factor~\cite{Bodey2019}. To extend spin coherence during resource-state generation, a Hahn-echo sequence must be implemented (this is inherent to the time-bin protocol). The inter-pulse delay $\tau_{\text{echo}}$ is chosen where the Hahn-echo visibility is maximised, while being compatible with timing restrictions of the generation sequence (Sec.~\ref{sec:classical}).
    \item Nuclear spin cooling~\cite{Jackson2022} before each round of resource-state generation extends the intrinsic electron spin dephasing time $T^*_2$. This reduces nuclear-induced spin decoherence during spin rotations by converging the randomly-spread nuclear spin states into a state with well-defined nuclear polarisation.
    \item For spin initialisation, it is advantageous to drive the non-cycling transition (e.g., $\ket{\downarrow}_x \to \ket{\Uparrow\uparrow\downarrow}_x$) which results in Raman scattering from the nuclear-dressed state $\ket{\downarrow,I_z}$ to $\ket{\uparrow,I_z}$ as required in nuclear spin cooling~\cite{Gangloff2019}. 
    \item The EPS unit emits photons encoded in time bins. To achieve this, a set of time-bin excitation pulses is generated by sending a laser pulse through an unbalanced interferometer with time delay $\tau_{\text{int}}$ (depicted in Fig.~\ref{fig:rsg}a), or picking two consecutive pulses from a pulsed laser~\cite{meng2023photonicfusionentangledresource}. 
\end{itemize}

Different types of resource states for FBQC can be generated with a single charged QD, including Greenberger–Horne–Zeilinger (GHZ) states, linear cluster states~\cite{lindner_proposal_2009}, and redundantly encoded cluster states, more generally called \textit{caterpillar} states. The EPS unit is highly reconfigurable since the geometry of the resource states only depends on the arrangement of spin rotation pulses, which are controlled by electronics. 

The underlying principle governing resource-state generation is the deterministic creation of entanglement via a waveguide-mediated spin-photon interface~\cite{Appel2021,appel_entangling_2022,Meng2024}. A spin-photon interface refers to a coherent coupling between a photon and the spin qubit, which in our case is realised by embedding a QD within a photonic-crystal waveguide. The planar waveguide serves a dual purpose: it facilitates efficient outcoupling of emitted photons and enables direct control of spin transitions, either through optical excitation from above or via microwave pulses delivered by a planar antenna. This configuration enables the implementation of pulse sequences that alternate between optical excitations and spin rotations, generating spin–photon entanglement. For instance, a spin-photon Bell state is prepared by:

\begin{enumerate}
    \item Applying a spin $\pi/2$-rotation pulse to an electron spin initialised in $\ket{\uparrow}_x$ to prepare a superposition state: $(\ket{\uparrow}_x+\ket{\downarrow}_x)/\sqrt{2}$.
    \item Exciting the cycling transition $\ket{\downarrow}_x\to\ket{\Downarrow\uparrow\downarrow}_x$ with an early pulse, resulting in $(\ket{\uparrow}_x \ket{\emptyset}+\ket{\downarrow}_x \ket{e})/\sqrt{2}$, where $\ket{\emptyset}$ is the vacuum state as $\ket{\uparrow}_x$ is not excited.
    \item Applying a spin $\pi$-rotation pulse to invert the correlations: $(\ket{\downarrow}_x \ket{\emptyset}-\ket{\uparrow}_x \ket{e})/\sqrt{2}$.
    \item Exciting the cycling transition $\ket{\downarrow}_x\to\ket{\Downarrow\uparrow\downarrow}_x$ again with a late pulse, leading to $\ket{\psi_{\text{Bell}}}=(\ket{\downarrow}_x \ket{l}-\ket{\uparrow}_x \ket{e})/\sqrt{2}$.
\end{enumerate}
$\ket{\psi_{\text{Bell}}}$ is a maximally entangled state between an electron spin and a time-bin encoded photon, where the photon emission time is correlated with the electron spin states. By iterating this protocol $M-1$ times, a time-bin encoded photon generated in each physical cycle is appended to the state, forming an $M$-qubit GHZ state~\cite{tiurev_fidelity_2021}.

Fig.~\ref{fig:rsg}b shows three types of resource states that can be generated using the above principle. The left panel shows the respective quantum circuits, each with a spin and $M$ ancillary photons initialised in $\ket{0}$. 
The right panel shows experimental pulse sequences to realise these protocols in the time-bin encoding. Hadamard gates are performed by applying spin $\pi/2$ pulses, whereas $C_{\mathrm{NOT}}$ gates are achieved by a pair of time-bin excitation pulses interleaved with a spin $\pi$-pulse~\cite{tiurev_fidelity_2021}. A Hahn-echo sequence must always be present, and therefore it is essential to maintain the same inter-pulse delay $\tau_{\text{echo}}$ throughout.

The redundantly encoded cluster state is the resource state required for the sFFCC architecture~\cite{Chan2025}. It resembles a linear cluster state with $M$ encoded qubits --- each of them is an $N$-photon GHZ state. This redundancy enables fusion between encoded qubits, where each encoded fusion has maximum $N$ attempts to perform a physical fusion between two resource states. 

In the following, we describe the physical components of the fusion measurement circuit. The entangled photons produced from all EPS units couple through an optical path or fibre to one fusion measurement circuit. The objective of the circuit is to route and measure these photons.

As illustrated in Fig.~\ref{fig:overview}a, its physical layout replicates the honeycomb lattice design of the sFFCC architecture. Although we consider a modular architecture where the EPS units and fusion measurement circuit reside on separate chips, monolithic integration onto a single chip is also possible. Here, each EPS unit outcouples photons to pass through a free-space unitary gate and into an optical fibre, before coupling into one fusion measurement chip through a fibre array. On this chip, photons are routed to specific fusion gates (blue squares) via a three-way photon switch (yellow triangles).

To implement the sFFCC fusion network (Fig.~\ref{fig:fusion_network}), the switching configuration cycles through three paths over time, as depicted in Fig.~\ref{fig:overview}c. The boundaries of the finite lattice change dynamically with time and require alternating single-qubit $X$-basis and $Z$-basis measurements at different layers~\cite{anyon_condensation_benbrown}, where only a single photon is routed to each fusion gate. As a result, each fusion gate at the boundary can simply be replaced by a photodetector.

\begin{figure}[hbtp!]
    \centering
	\includegraphics[width=1\linewidth, trim=0.cm 0cm 0.cm 0.0cm,clip]{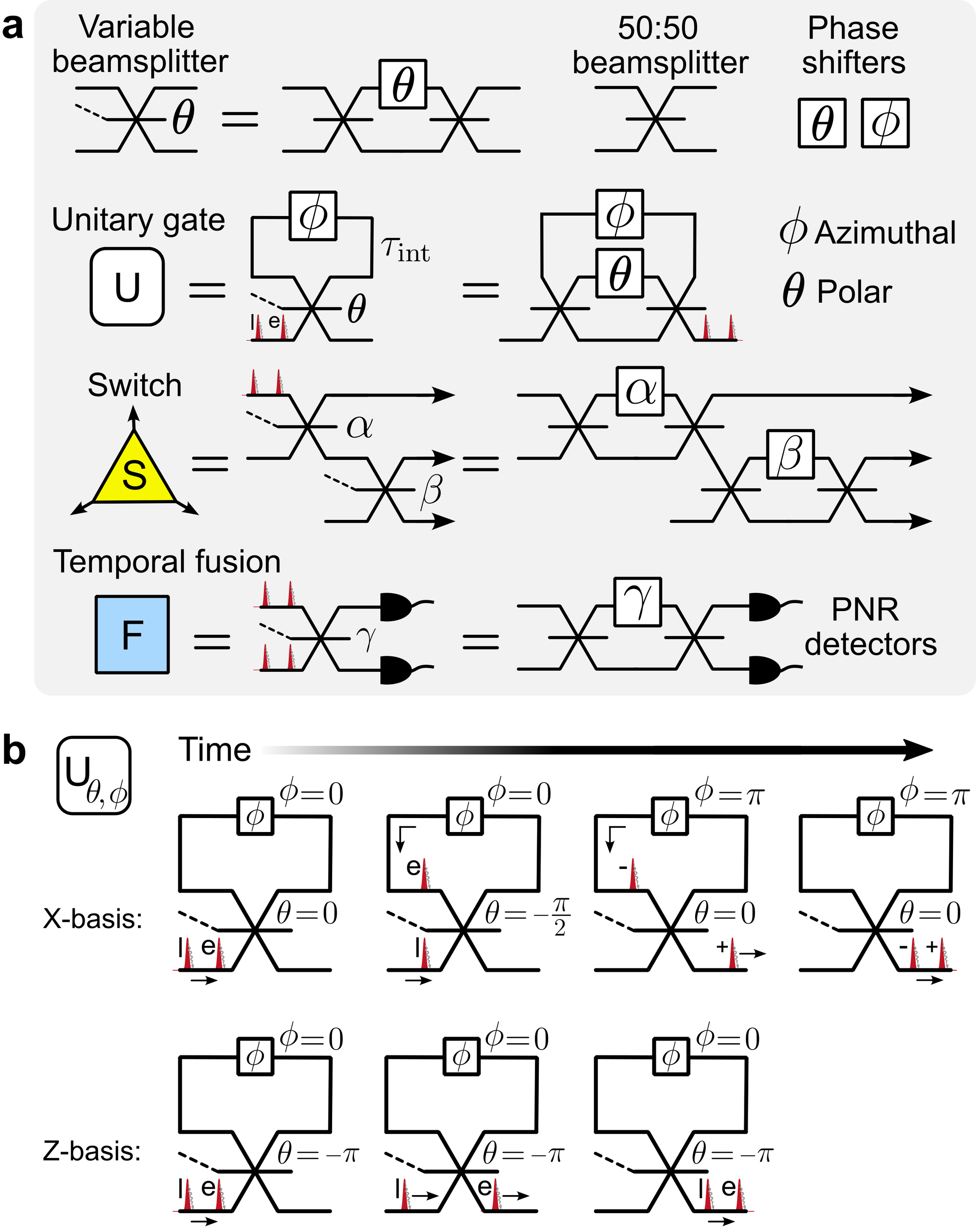}
    \caption{\textbf{Hardware components of a fusion measurement circuit.} \textbf{(a)} A variable beamsplitter (VBS) with controllable reflectivity is central to all physical components on the fusion measurement circuit. It consists of an actively controlled phase shifter and two passive 50:50 beamsplitters. The unitary gate applies either a Hadamard gate ($X/Y$-basis) or an identity operation ($Z$-basis). The three-way switch operates by adaptively changing the VBS reflectivity. Temporal fusion  requires only two photon-number resolving (PNR) detectors. \textbf{(b)} Operational principle of a time-bin encoded unitary gate for $X$-basis and $Z$-basis projection. For the $X$-basis, the early time-bin is swapped into a delay loop to interfere with the late time-bin. For the latter, both time bins bypass the delay path.
    }
    \label{fig:component}
\end{figure}

Now, we describe the operational principles of all the optical elements required for the FBQC architecture of Fig.~\ref{fig:overview}b. At the core of all components is a variable beamsplitter (VBS), the reflectivity of which is actively controlled. A VBS is realised using two passive 50:50 beamsplitters and one active phase shifter which imprints a relative phase of $\theta$ between two optical paths (Fig.~\ref{fig:component}a). Essentially, it is a Mach–Zehnder interferometer with unitary transformation for two input paths:
\begin{align}
    \mathcal{U}_{\text{VBS}} \sim \begin{bmatrix}
    \sin{(\theta/2)} & \cos{(\theta/2)}\\
    \cos{(\theta/2)} & -\sin{(\theta/2)}
    \end{bmatrix},
\end{align}
which, up to a global phase, represents a beamsplitter with varying reflectivity $\sin^2(\theta/2)$.
The unitary gate performs a single-qubit rotation on the time-bin encoded photons, depending on the fusion measurement basis. Since the entangled photons emitted from an EPS unit are represented in the $Z$-basis ($\ket{e}$ and $\ket{l}$), the default unitary gate is a Hadamard gate to rotate them onto the $X$-basis, as required in $\{ZZ,XX\}$ physical fusions. For single-qubit measurements in biased fusion, however, the unitary gate needs to be reconfigured to perform an identity operation. In the time-bin picture, the unitary gate is realised by connecting one output path with one input of a VBS (Fig.~\ref{fig:component}a), introducing a time delay $\tau_{\text{int}}$. In essence, it consists of two adaptive phase shifters and two beamsplitters, acting as a time-bin interferometer. A phase shifter in the VBS controls the polar angle $\theta$ on a Bloch sphere, while the other phase shifter in the delay path sets the azimuthal angle $\phi$.

Figure~\ref{fig:component}b illustrates the time evolution of a time-bin encoded photon passing through a unitary gate that implements either a Hadamard (for $X/Y$-basis measurements) or identity ($Z$-basis) operation. For $X$-basis projection, the early time-bin is first swapped into a delay path by setting $\theta=0$. Prior to the interference of two time bins at the VBS, we set the phase to $\theta=-\pi/2$ such that the $\ket{+} = (\ket{e} + \ket{l})/\sqrt{2}$ component exits through the alternate output port of the VBS, while the $\ket{-} = (\ket{e} - \ket{l})/\sqrt{2}$ component enters the delay loop and acquires a phase shift with $\phi=\pi$. Resetting $\theta = 0$ then releases the $\ket{-}$ component. Modulating $\phi=\pi/2$ while the early time-bin resides in the delay loop induces a relative phase shift between the early and late bins, rotating to $Y$-basis. For the identity operation, $\theta = -\pi$ is maintained throughout, causing both time bins to be reflected at the VBS and bypass the delay loop. In Fig.~\ref{fig:overview}b, while the unitary gates serve only to control the fusion failure basis before the fusion gates, positioning them upstream instead of downstream of the photon switch is more resource-efficient, as this eliminates the need for a unitary gate per path after the switch. Furthermore, given that the time-bin interferometer within the unitary gate must be actively stabilised, and the delay loop must be long enough to match $\tau_{\text{int}}$, a free-space implementation of the unitary gate may be preferable to the on-chip alternative.

The three-way photon switch can simply be realised by two stacked VBSs. In general, when a photon needs to be routed to the first optical path, the first VBS is set to fully reflect both time bins. For the second path, the first VBS transmits while the second reflects. Note that photons on all three paths must take the same amount of time to arrive at the respective fusion gates. 

Since we consider temporal fusion, the fusion gate (Fig.~\ref{fig:component}a) uses only two photodetectors and one VBS which switches between two-photon fusion measurements ($\gamma=-\pi/2$) and single-qubit measurements ($\gamma=-\pi$). Alternatively, the fusion gate can be made fully passive by replacing the VBS with a 50:50 beamsplitter and leveraging the reconfigurability of excitation pulses in the EPS units. This strategy relies on \textit{excitation-based feedback}, as detailed in Sec.~\ref{resource_scaling}. The photons are then measured by on-chip single-photon detectors, which must be highly efficient, photon-number resolving (to distinguish between fusion failure and erasure due to loss) and have low dark counts. The photon counts are converted into time tags via an external time-to-digital converter, and subsequently analysed by a classical control unit to interpret the fusion measurement outcome. 

The physical chip consisting of photon switches and fusion gates could be fabricated on low-loss material platforms, for example, silicon nitride (SiN)~\cite{Alexander2024} or single-crystal thin lithium niobate [LiNbO$_3$ (LN)] films bonded on a silica insulating substrate [lithium niobate on insulator (LNOI)]~\cite{Wang2018}. In both cases, active phase shifters would need to be implemented electro-optically on the LNOI platform and can be realised, e.g., via heterogeneous integration on SiN. Both platforms are fully compatible with QD photon sources where pilot demonstrations have already been reported~\cite{Sund2023,Wang2023}. 
QD single-photon and entangled-photon sources emitting in the telecom band~\cite{albrechtsen2025,hauser2025deterministichighlyindistinguishablesingle,laccotripes2025entangledphotonsourcetelecom} have also been realised, eliminating the need for frequency conversion.

\subsection{Classical control and feedback}
\label{sec:classical}

\begin{figure*}[]
	\includegraphics[width=0.9\linewidth, trim=0.cm 0.05cm 0.cm 0.0cm,clip]{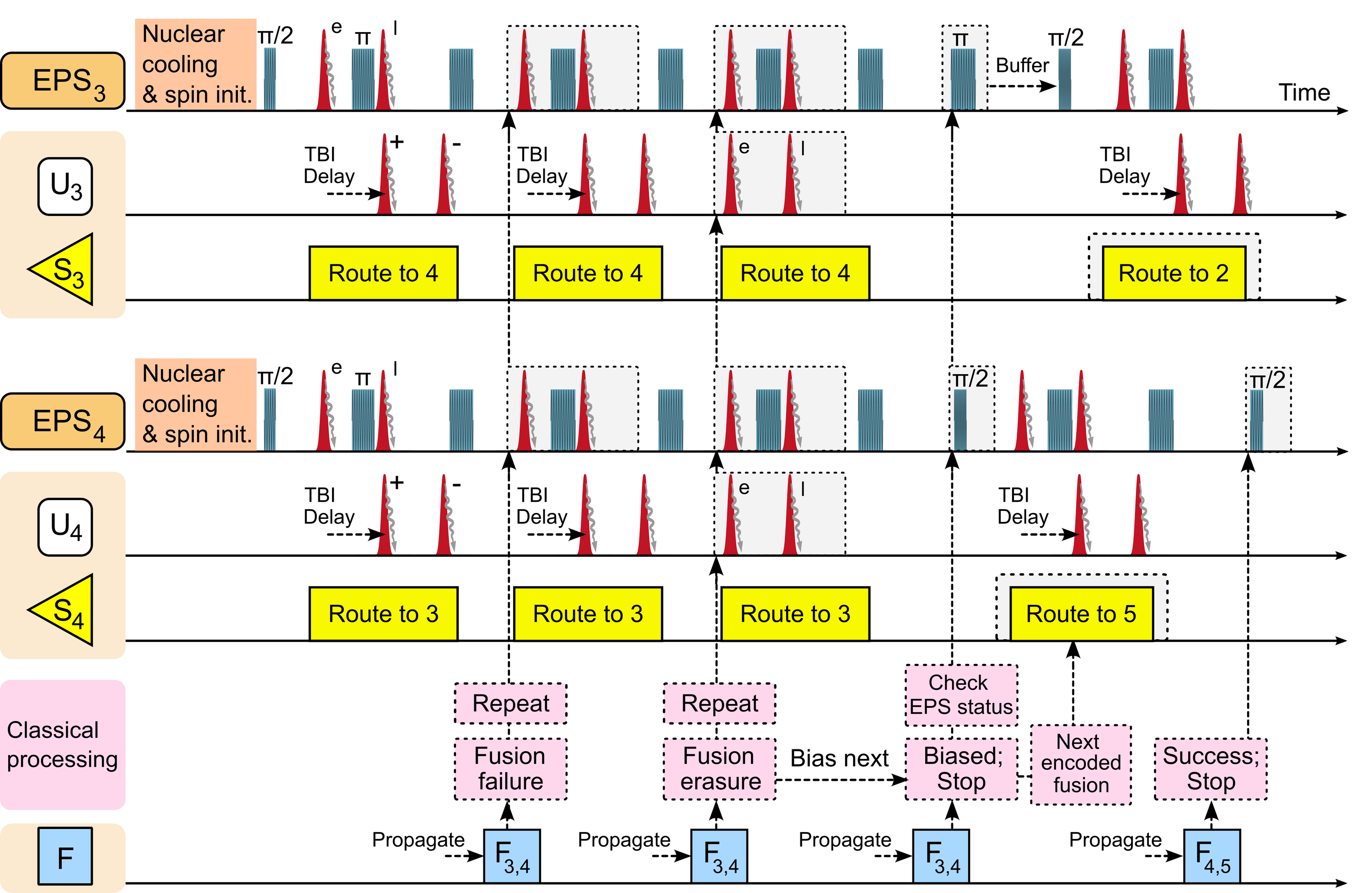}
    \centering
    \caption{\textbf{Snippet of the experimental control sequence}. Central to the control sequence is the generation protocol of a redundantly encoded linear cluster state with two encoded qubits of different code sizes. Each EPS sends a time-bin encoded photon to a unitary gate which rotates it onto either the $X$- or $Z$- basis. For $X$-basis, the time bins are delayed by $\tau_{\text{int}}$ (labelled as TBI delay). The propagation delay of a photon from the EPS to a fusion gate is included. To illustrate all possible cases, we consider a scenario where EPS$_3$ and EPS$_4$ attempt an encoded fusion $F_{3,4}$: the first physical fusion results in a failure, the second attempt leads to erasure due to photon loss, and the third attempt is biased in $ZZ$. The encoded fusion $F_{4,5}$ between EPS$_4$ and EPS$_5$ (pulse sequence not shown) is successful on the first attempt. A buffer time of $\tau_{\text{echo}}$ is added to EPS$_3$ to synchronise photon generation with EPS$_2$.}
    \label{fig:control}
\end{figure*}

A classical control unit is comprised of devices that process photon detection events from the fusion measurement circuit, interpret the corresponding fusion outcomes, and relay control commands to both the EPS units and the respective components on the fusion measurement circuit (Fig.~\ref{fig:overview}b). The communication infrastructure may include a service layer responsible for distributing a master clock signal to all relevant equipment~\cite{Pittaluga2025}, such as arbitrary waveform generators, field-programmable gate arrays, and time-to-digital converters at the EPS units located on the management layer, enabling a system-wide clock synchronisation.

With a clear understanding of the physical hardware, we now explain the workflow logic of the classical control unit to run an operation in a quantum algorithm. An algorithm consists of a series of quantum operations to be performed on the physical hardware, where each operation is implemented by a combination of fusion measurements in different bases~\cite{bartolucci_fusion-based_2023}. 

We define the time required to complete all encoded fusions within a single layer of the sFFCC lattice (Fig.~\ref{fig:overview}c) as one \emph{physical clock cycle}. The total time needed to complete all layers up to code distance $L$ then defines one \emph{logical clock cycle}. For each layer, the EPS-emitted photons are routed to their respective fusion gates to perform RUS encoded fusions. For each encoded fusion, a maximum number of $N$ physical fusions are attempted. The measurement outcomes are obtained from photon detection and interpreted according to the click patterns in Fig.~\ref{fig:type}, which can result in one of two scenarios:
\begin{itemize}
    \item Repeat: If a physical fusion fails or has photon loss, the corresponding EPS units emit a new pair of photons while the unitary gates apply either a Hadamard or identity operation, respectively. 
    \item Next encoded fusion: Once either or both encoded fusion outcomes are obtained, or $N$ is reached, the workflow advances to the next layer. A status check is then performed to ensure all emitters on the current layer are ready: If only a subset of encoded fusions on a layer is completed, the corresponding emitters are idled until the remaining fusions on that layer are finished. Once the check is passed, the configurations for EPS units, unitary gates, and photon switches are updated accordingly for the next encoded fusion.
\end{itemize}
After completing all encoded fusions, their measurement outcomes are processed to construct a syndrome graph, which is passed to a classical decoder for error correction. The corrections do not need to be applied directly to the quantum system, but are tracked with the so-called Pauli frame~\cite{bartolucci_fusion-based_2023}, which is a list of Pauli operators that account for any errors accumulated during the logical clock cycle. The Pauli frame is consistently updated after every operation and is eventually applied to the classical outcomes to correctly interpret the logical qubit state.

Following the discussion of the system hardware and operational logic, we now present a snippet of the experimental pulse sequence in real time to elucidate the synergistic operation of various components. Specifically, we consider the pulse sequences for two EPS units (EPS$_3$ and EPS$_4$), the associated unitary gates, switches, and fusion gates, as presented in Fig.~\ref{fig:control}. We assume the use of RUS encoded fusions, with physical temporal fusions.

First, EPS$_3$ and EPS$_4$ initialise their spin states with nuclear cooling and spin pumping pulses. Reading out the spin effectively clears all memory from the previous iteration. From here, the spin-photon entanglement sequence begins. A time-bin encoded photon is generated from each EPS unit. The two photons are rotated onto $X$-basis via the unitary gates, and are subsequently routed to the same fusion gate. 

For pedagogical purposes, here we consider four different fusion outcomes which capture all possible cases: The first attempt of physical fusion $F_{3,4}$ has resulted in fusion failure, thus sending a ``Repeat" command to both EPS units (dashed line) which then emit another pair of photons. The second attempt has led to fusion erasure due to photon loss, therefore the classical control unit commands both EPS units to emit a new pair of photons and their unitary gates to perform an identity operation for $Z$-basis rotation, then reconfigures the fusion gate $F_{3,4}$ to perform single-qubit measurements. As a result, on the third attempt, biased fusion is performed which has successfully recovered the $\overline{ZZ}$ operator. A status check then reveals that EPS$_2$ (EPS$_5$) is not ready (ready), thus EPS$_3$ (EPS$_4$) applies a spin $\pi$- ($\pi/2$-) pulse. The additional spin $\pi$-pulse not only introduces a buffer time of $\tau_{\text{echo}}$, but also acts as a refocusing pulse in the Hahn echo which is already built into the generation sequence. This means the emitter in EPS$_3$ can be idled with minimal degradation of coherence. On the other hand, a $\pi/2$-pulse prepares the spin before generating another encoded qubit in the linear cluster state (Recall that the primitive sFFCC resource state is a redundantly encoded linear cluster state, in which fusions are performed on the encoded qubits. The resource-state generation sequence includes alternating spin $\pi/2$ and $2N$ $\pi$-pulses, where a $\pi/2$ pulse is applied before generating the $N$-photon encoded qubit). 
Intriguingly, it also terminates the Hahn echo by projecting the spin onto its ground state, thereby preventing further decoherence. At last, the first fusion attempt $F_{4,5}$ succeeds, recovering both $\overline{XX}=XX$ and $\overline{ZZ}=ZZ$.

For any physical fusion, if one or both photons involved are lost, the spin(s) entangled with the lost photon(s) effectively become a mixed state. In the case of RUS gates in spin-optical architectures~\cite{hilaire2024enhancedfaulttolerancephotonicquantum,deGliniasty2024spinopticalquantum}, this corresponds to a spin phase scramble applied to the emitter, and an appropriate spin gate must be applied as a correction. This requires fast feedback between identifying the detection pattern and implementing spin control. 

However, with the sFFCC, this feedback is not required: even if the spin is a mixed state, the same pair of excitation pulses can still be performed on the spin to generate another time-bin encoded photon with a resultant state $\rho_{\text{mixed}}=\ket{e}\ket{\uparrow}\bra{\uparrow}\bra{e}+\ket{l}\ket{\downarrow}\bra{\downarrow}\bra{l}$. The photon emitted will not be entangled with the spin but the information on their $Z$-correlation is preserved. This is fully compatible with the biased physical fusion in the following attempt, where only single-qubit $Z$ measurement is performed on the photon. 

It is worth noting that biased fusion is equivalent to indirect spin initialisation: As $\rho_{\text{mixed}}$ is stabilised by the $ZZ$ operator, measuring the photon in the $Z$-basis projects the spin onto the same basis. Furthermore, the Hahn-echo sequence is effectively restarted, since prior spin-photon correlations are erased. In other words, photon loss could be leveraged with biased fusion to re-initialise both the spin state and Hahn echo. This is experimentally favourable, as the number of required spin control pulses is reduced, thereby minimizing error accumulation from pulse imperfections.

The experimental sequence in Fig.~\ref{fig:control} requires electronic and optical pulses to fulfil specific timing constraints. These constraints limit the maximum allowable feedback time and impose a minimum speed requirement on phase shifters used to implement RUS fusions.

\begin{figure}[hbtp!]
    \centering
	\includegraphics[width=1\linewidth, trim=0.cm 0cm 0.cm 0.0cm,clip]{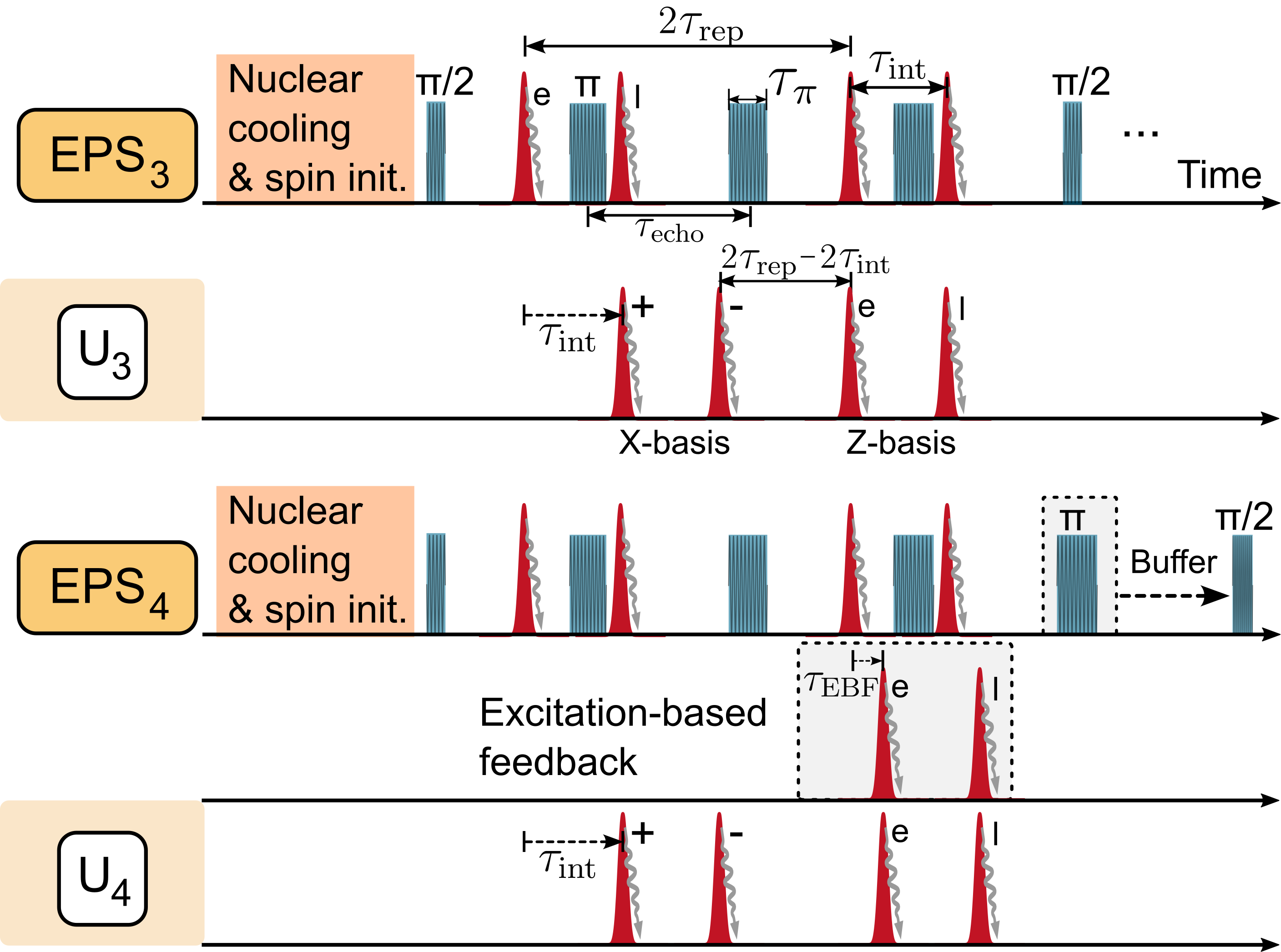}
    \caption{\textbf{Timing restrictions.} We consider the pulse sequences for two EPS units, with the relevant timing offsets. A buffer pulse is added to EPS$_4$ to synchronise with another EPS unit (not shown). For EPS$_4$, the addition of a fibre-based electro-optic modulator and a short delay loop on its excitation path delays both time bins by $\tau_{\text{EBF}}$. This removes the temporal overlap between photons from two EPS units, enabling single-qubit measurements without reconfiguring the fusion gate. The timing requirement between different pulses is derived and explained in the main text.}
    \label{fig:timing}
\end{figure}

Fig.~\ref{fig:timing} showcases an example of a pulse sequence to perform a RUS encoded fusion between EPS$_3$ and EPS$_4$. From here, a set of inequalities is determined:
\begin{align}
    \tau_{\text{rep}} &= \tau_{\text{echo}};\label{eq:ineq1}\\ 
    \tau_d &< \tau_\text{int}; \label{eq:ineq2}\\ 
    \tau_p &<2 \tau_{\text{rep}} - 2\tau_{\text{int}};\label{eq:ineq3}\\ 
    \tau_\pi &<\tau_{\text{int}}<\tau_\text{echo};\label{eq:ineq4}\\ 
    \tau_{\text{TB}}&<\tau_{\text{EBF}}<\tau_{\text{int}}-\tau_\pi;\label{eq:ineq_EBF}\\ 
    \tau_{\text{PS}}&<\text{min}\{\tau_{\text{int}},2 \tau_{\text{rep}} - 2\tau_{\text{int}}\},\label{eq:ineq5} 
\end{align}
where $\tau_{\text{rep}}$ is the repetition period of the excitation laser, $\tau_{\text{echo}}$ is the time separation between consecutive spin control pulses in which the spin-echo visibility $V_{\text{echo}}$ is maximised~\cite{Meng2024}, $\tau_{d}$ is the photodetector deadtime, $\tau_{\text{int}}$ is the interferometric delay between early and late time bins in the unitary gate, $\tau_p$ is the feedforward processing time, $\tau_\pi$ is the spin $\pi$-pulse duration, $\tau_{\text{TB}}$ is the duration of a time-bin pulse and $\tau_{\text{EBF}}$ is the time delay required for excitation-based feedback (see Sec.~\ref{resource_scaling}), and $\tau_{\text{PS}}$ is the time required for an active phase shifter to switch between $\theta=0$ and $\theta=\pi$. In the following paragraphs, we explain the rationale behind each inequality:
\begin{itemize}
    \item Each EPS unit by default generates time-bin photons every $2\tau_{\text{rep}}$. To avoid overlap between optical excitation and spin refocusing pulses, it is necessary that $\tau_{\text{rep}}=\tau_{\text{echo}}$ (Eq.~(\ref{eq:ineq1})). If all emitters on the same layer need to synchronise before the next encoded fusion, buffer spin-control $\pi$-pulses can be applied to introduce a delay. After adding one buffer pulse (Fig.~\ref{fig:timing}), the next pair of excitation pulses will appear after $3\tau_{\text{rep}}$, thus requiring the clock period of the excitation laser to be $\tau_{\text{rep}}$, which enables pulse-picking of the laser at odd units of $\tau_{\text{rep}}$.
    \item Temporal fusion conditions on detecting two-fold coincidence clicks between early and late time bins; therefore the photodetector deadtime $\tau_d$ must be shorter than $\tau_{\text{int}}$ (Eq.~(\ref{eq:ineq2})). Additionally, in the case of biased fusion, the minimum allowable time separation between consecutive optical excitation pulses is $2\tau_{\text{rep}}-2\tau_{\text{int}}$ (see unitary gate $U_3$ in Fig.~\ref{fig:timing}). This constraint places an upper bound on the allowable feedforward time since the previous physical fusion needs to be processed and the next pair of excitation pulses must be sent. For brevity, we define the total time $\tau_p$ taken for (i) photons to travel from unitary gates to detectors, (ii) retrieving and processing of the fusion outcome, and (iii) applying the next excitation pulse, which needs to be shorter than $2\tau_{\text{rep}}-2\tau_{\text{int}}$ (Eq.~(\ref{eq:ineq3})). 
    \item $\tau_{\text{echo}}$ cannot be too short since a spin $\pi$-pulse must be sandwiched between early and late time bins separated by $\tau_{\text{int}}$. For the same reason, $\tau_{\text{int}}$ must be longer than the spin $\pi$-pulse duration $\tau_\pi$ to prevent pulse overlap (Eq.~(\ref{eq:ineq4})).
    \item The minimum switching time of the phase shifter $\tau_{\text{PS}}$ in a unitary gate must be shorter than the interferometer delay $\tau_{\text{int}}$. In Fig.~\ref{fig:component}b, for $X$-basis projection, a $\pi$ phase shift must be applied before a photon exiting the delay loop. If there is a failure-basis change in the unitary gate due to biased fusion, as shown in Fig.~\ref{fig:timing}, $\tau_{\text{PS}}$ should also be shorter than $2 \tau_{\text{rep}} - 2\tau_{\text{int}}$ (Eq.~(\ref{eq:ineq5})). 
\end{itemize}

To provide concrete numbers, Table~\ref{tab:FOM} lists the relevant system parameters (P) alongside the corresponding values reported for existing quantum dot platforms. This enables a direct comparison between state-of-the-art capabilities and the timing requirements outlined in our architecture. For instance, feedforward that combines single-photon detection with both electronic logic processing and optical modulation has been demonstrated at cryogenic temperatures with a latency of $23$~ns~\cite{Thiele2025}, which is already on the order of $< 2\tau_{\text{echo}} - 2\tau_{\text{int}} \approx 34.3$~ns.

\begin{table}[hbtp!]
  \begin{center}
    \begin{tabular}{|c|c|c|}
      \hline
      \textbf{P} & \textbf{Description} & \textbf{Reference value} \\ 
      
        \hline
        $\tau_{\text{int}}$ & Time-bin interferometer delay & $11.83$~ns~\cite{Meng2024} \\
        $\tau_{d}$ & Detector dead time & $\sim10$~ns \\
        $\tau_{\text{echo}}$ & Echo rephasing time & $29$~ns~\cite{Meng2024}; $75$~ns~\cite{hogg2024fastopticalcontrolcoherent} \\
        $\tau_{\pi}$ & spin $\pi$-pulse duration & $4$~ns~\cite{Meng2024} \\
        $\tau_{\text{rep}}$ & Pulse repetition period & $13.8$~ns~\cite{Meng2024} \\
                          \hline         
      
    \end{tabular}
    \end{center} 
 \caption{\textbf{System parameters and reference values.} It should be noted that these values currently do not satisfy all timing inequalities, owing to equipment constraints~\cite{ChanThesis}; they are provided solely as indicative timescales.}
    \label{tab:FOM}
\end{table}


\subsection{General error models and performance}
\label{sec:errormodels}

\begin{figure*}[hbtp!]
  \centering
  \includegraphics[width=\textwidth]{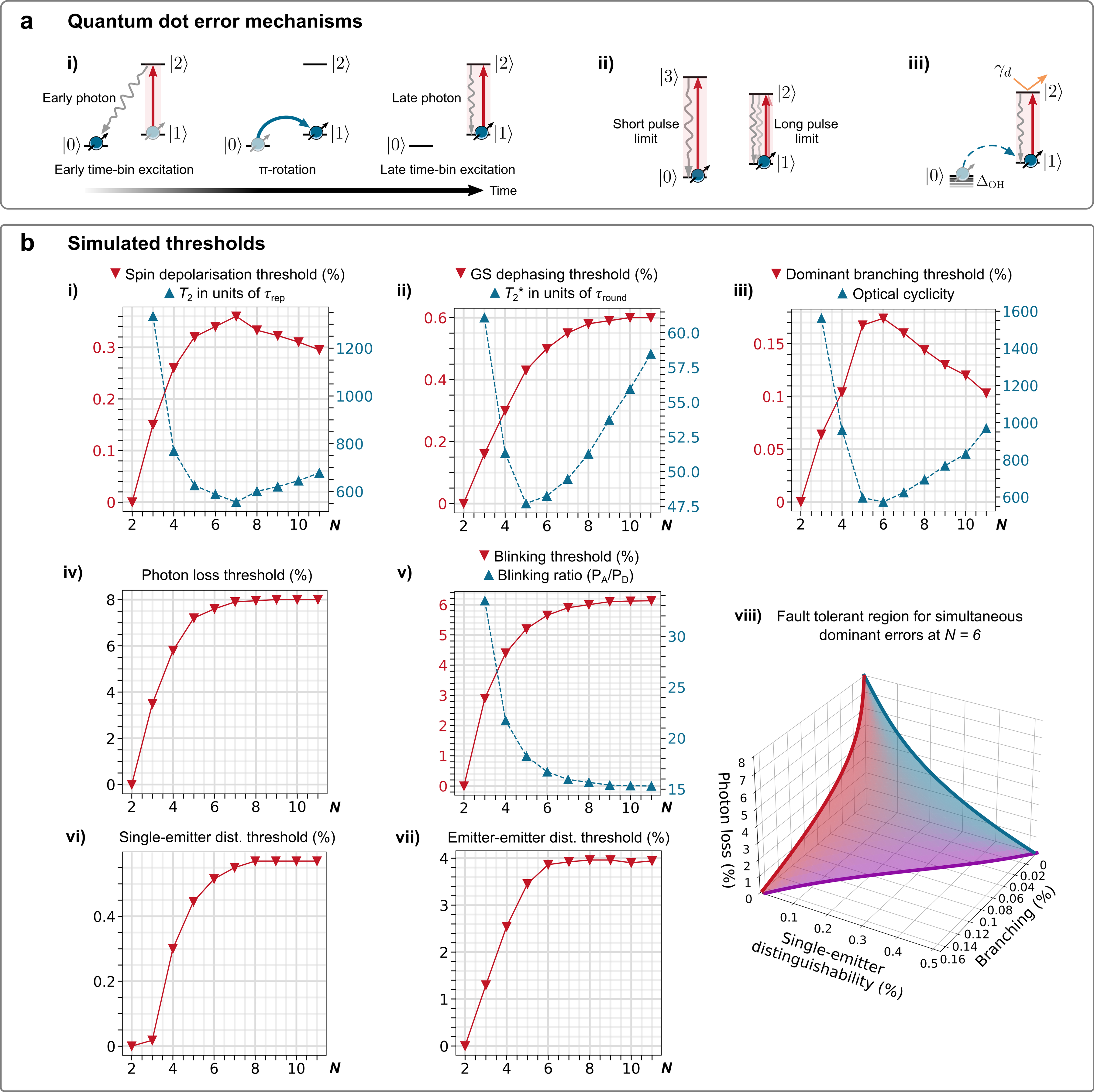}
  \caption{\textbf{Threshold simulations under all relevant realistic QD error sources}. \textbf{(a)} Quantum dot error mechanisms in the time-bin picture. The red highlighted $\ket{1}\leftrightarrow\ket{2}$ transition is optically excited to produce a photon. In an ideal emission round, one photon from the early or late time-bin is emitted through the same transition. i) Dominant branching errors: an early photon is emitted via a diagonal (branched) decay path, followed by a correctly-emitted late photon. ii) Excitation errors: short pulses may off-resonantly drive $\ket{0}\leftrightarrow\ket{3}$. Long pulses may cause two-photon emission via $\ket{1}\leftrightarrow\ket{2}$. iii) Interaction with the nuclear spin bath causes ground state dephasing via Overhauser field fluctuations $\Delta_{\text{OH}}$. Laser-induced spin-flip errors occur during spin rotation (dashed blue). Scattered phonons induce a random phase change with rate $\gamma_d$ (orange). \textbf{(b)} Thresholds for increasing $N$. i) Spin depolarising noise (spin $X/Z/Y$) during the time-bin protocol is optimally tolerated at 0.36\% for $N = 7$, decreasing thereafter. ii) Ground state dephasing (as depicted in a(iii)) can be approximated as Markovian spin $Z$ errors in the low-error limit, saturating at a 0.6\% threshold after $N = 8$. The corresponding $T^*_2$ requirements for protocols without spin echo (in units of time taken to produce one physical photon, $\tau_{\text{round}}$) for each threshold point is shown. iii) The dominant branching error term (as depicted in a(i)) causes simultaneous spin-photon X errors that propagate to the next logical fusion, with a threshold that eventually decreases after a peak at $N = 6$. From this, the corresponding optical cyclicity target for each threshold value is obtained. iv) Photon loss tolerance for all loss sources saturates within $N = 8$ physical fusions. v) Correlated loss from blinking, determined by the ratio of the dead-to-alive and alive-to-dead rates ($P_A/P_D$) of the quantum dot. Higher $N$ provides higher tolerance. Performance is reduced compared to uncorrelated loss. vi) Single-emitter distinguishability, captured by photon $Z$ errors, achieves 0.57\% tolerance. vii) Emitter-emitter indistinguishability, where partial distinguishability between photons from different emitters reduces the Hong-Ou-Mandel visibility $V^{ee}_{\text{HOM}}$, is tolerated up to a 4\% reduction.  viii) Fault-tolerant region for simultaneous photon loss, branching, and single-emitter distinguishability, for $N = 6$. 
} 
  \label{fig:all_err_stuff}
\end{figure*}

Efficient quantum error correction requires a thorough understanding of the underlying error mechanisms present in any physical hardware in order to achieve reliable, fault-tolerant computation. Correlated and biased physical errors are of particular significance, since thresholds are usually derived assuming Markovian noise and non-Markovian error processes can shift these thresholds significantly. Additionally, a hardware-tailored decoder could exploit such information to achieve higher thresholds in a practical setting~\cite{NickersonBrown2019}. 

In this section, we analyse the performance of the sFFCC architecture under experimentally relevant error sources in QD systems. While models of general loss, circuit-level spin errors, and distinguishability errors during resource-state generation according to the Lindner-Rudolph protocol~\cite{lindner_proposal_2009} have been investigated for the sFFCC lattice~\cite{Chan2025}, intrinsic physical errors encountered in QD-based resource-state generation have hitherto been largely unexplored in terms of their impacts on fault-tolerant thresholds.

To accurately reflect the impact of real-world errors inherent to experimental hardware, we extend the previous analysis in two key ways: first, we consider the time-bin resource-state generation protocol described in Sec.~\ref{architecture}. Second, we examine a full catalogue of physical noise sources intrinsic to QDs, and map their effects onto circuit-level errors impacting fusion gates. This enables us to derive more realistic fault-tolerant thresholds and translate them into concrete hardware performance targets that serve as practical experimental benchmarks. 

The physical noise sources we consider include branching error due to finite optical cyclicity, spin-flip errors associated with spin rotation pulses, optical excitation errors, emitter-emitter distinguishability due to slow spectral diffusion, fast phonon-induced pure dephasing leading to single-emitter distinguishability, spin depolarising error (finite $T_2$), emitter blinking, and ground-state dephasing. This constitutes an exhaustive catalogue of error processes that are essential for real-world implementations. Quantum dot sources are becoming increasingly mature as their dominant physical error channels have already been identified and understood. This foundational understanding is a critical prerequisite for any physical quantum computing platform before transitioning to the engineering phase, where efforts shift toward mitigating these errors and scaling up the technology.

We show that all these noise sources can be simplified as either photon loss, or Pauli errors that permeate through resource-state generation, resulting in imperfect photonic resource states. Spin-related noise sources, in particular, manifest as spin Pauli errors during the generation sequence, which are propagated to photons through photon emission ($C_{\mathrm{NOT}}$ gate) described by the following relations~\cite{Chan2025}:
\begin{align}
C_{\mathrm{NOT}}(X_{\text{spin}}\otimes I_{\text{photon}})\,C_{\mathrm{NOT}}^{-1}
&= X_{\text{spin}}\otimes X_{\text{photon}};\nonumber\\
C_{\mathrm{NOT}}(Z_{\text{spin}}\otimes I_{\text{photon}})\,C_{\mathrm{NOT}}^{-1}
&= I_{\text{spin}}\otimes Z_{\text{photon}},
\label{eq:cnot_map}
\end{align}
where the spin (control) and each photon (target) are initialised in $\ket{+}$ and $\ket{0}$, respectively. $I$ is the identity gate. $X, Z$ are Pauli operators representing circuit errors.

We first consider branching error due to finite cyclicity.
As described in Sec.~\ref{architecture}, the time-bin protocol requires repeated excitation and photon emission through the same optical transition. In practice, the excited state may decay through the diagonal transition with a small probability due to finite cyclicity. This \textit{branching} error not only results in the emission of a photon with an undesired frequency, but also an incoherent spin flip. 

We analyse the scenario in which the excited state $\ket{2}$ decays erroneously via the non-cycling diagonal transition to $\ket{0}$ following an early time-bin excitation. Subsequently, after a spin $\pi$-pulse, a late photon is correctly emitted through the cycling transition, as depicted in Fig.~\ref{fig:all_err_stuff}a(i). This is equivalent to applying an $X$ error to both the spin and the emitted time-bin photon. While other branching mechanisms exist~\cite{Appel2021_phdthesis, tiurev_fidelity_2021}, they result in either photon loss or spin-flip-equivalent errors.

For the generation of encoded linear cluster states (Fig.~\ref{fig:rsg}b), a branching error that occurs in the $i$-th photon of an encoded qubit therefore manifests as both a spin $X$ error and an $X$ error on the $i$-th photon (If the error occurs for the very first photon emission, only the photon incurs an \(X\) error as there is obviously no previous spin-photon state with which to compare, and generation proceeds normally). From Eq.~(\ref{eq:cnot_map}), a $C_{\text{NOT}}$ gate ``copies" spin $X$ errors onto a photon, thereby spreading the error; spin $Z$ errors are merely transferred from the spin to the photon, localising the error. Consequently, the $X$ error on the spin is retained after affecting all subsequent photons of an encoded qubit. The Hadamard gate that generates the next encoded qubit converts this spin $X$ error into a spin $Z$ error, which is then localised to the first photon of the encoded qubit. Thus, branching can lead to a two-encoded fusion error, and thus we may expect to see a $2\times$ reduction of the peak threshold compared to a spin $Z$ error model (further decreased by the correlated nature of the error). 

Indeed, in Fig.~\ref{fig:all_err_stuff}b(iii), we observe a $\approx3\times$ reduction in threshold compared to that under the spin $Z$ error model in Fig.~\ref{fig:all_err_stuff}b(ii). The fault-tolerance threshold for branching peaks at $0.174\%$ for $N=6$ (Fig.~\ref{fig:all_err_stuff}b(iii)), then falls with $N$ as the probability of a branching event increases with more photon emissions. The effect of branching on the evolution of the resource state is analysed in \SM B. 

The branching probability $p_b$ is related to optical cyclicity $C$ via the following definitions~\cite{tiurev_fidelity_2021}:
\begin{align}
  C \equiv \frac{\Gamma_{\rm c}+\gamma_{\rm c}}{\Gamma_{\text{nc}}+\gamma_{\text{nc}}};\quad p_b \equiv \frac{\Gamma_{\text{nc}}}{\Gamma_{\rm c} + \Gamma_{\text{nc}}}
    <\frac{1}{C + 1},\label{eq:C}
\end{align}
where $\Gamma_{\rm c}$ ($\gamma_{\text{c}}$) is the radiative decay (loss) rate of the cycling transition, and $\Gamma_{\rm nc}$ ($\gamma_{\text{nc}}$) is the radiative decay (loss) rate of the non-cycling transition. The inequality in Eq.~(\ref{eq:C}) is valid when $\gamma_{\rm c}\approx\gamma_{\text{nc}}\ll1$.

This means that for a peak threshold of 0.174\% observed at $N=6$, the QD must cycle through at least $C=574$ correct transitions before a branching error or photon loss occurs.

\begin{table*}[hbtp!]
\centering
\renewcommand{\arraystretch}{1.2}
\small

\begin{tabular}{|l|l|c|l|l|}
\hline
\makecell{\textbf{Experimental}\\\textbf{error source}}
& \makecell{\textbf{Error}\\\textbf{model}}
& \textbf{\textit{N}}
& \makecell{\textbf{Maximum tolerable}\\\textbf{threshold}}
& \makecell{\textbf{Required experimental}\\\textbf{benchmark}} \\
\hline\hline

\makecell[l]{General loss}
& Uncorrelated loss
& 8
& \makecell[l]{8\% \\ {[Fig.~\ref{fig:all_err_stuff}b(iv)]}}
& $\eta > 92\%$ \\
\hline

\makecell[l]{Branching \\ (finite cyclicity)}
& \makecell[l]{Spin-photon $X$ errors}
& 6
& \makecell[l]{0.174\% \\ {[Fig.~\ref{fig:all_err_stuff}b(iii)]}}
& \textit{C} $> 574$ \\
\hline

\makecell[l]{Emitter-emitter distinguishability \\ (slow spectral diffusion)}
& Distinguishability
& 8
& \makecell[l]{4\% \\ {[Fig.~\ref{fig:all_err_stuff}b(vii)]}}
& $V^{ee}_{\text{HOM}} > 96\%$ \\
\hline

\makecell[l]{Single-emitter distinguishability \\ (fast pure phonon dephasing)}
& Photon $Z$ errors
& 8
& \makecell[l]{0.57\% \\ {[Fig.~\ref{fig:all_err_stuff}b(vi)]}}
& $V^{se}_{\text{HOM}} > 99.4\%$ \\
\hline

Optical excitation errors
& Distinguishability
& 8
& \makecell[l]{4\% \\ {[Fig.~\ref{fig:all_err_stuff}b(vii)]}}
& $V^{ee}_{\text{HOM}} > 96\%$ \\
\hline

\makecell[l]{Laser-induced spin-flips \\ during spin control}
& \makecell[l]{Loss and spin \\ depolarisation errors}
& 7
& \makecell[l]{0.12\% \\ {[Fig.~\ref{fig:all_err_stuff}b(i)/3]}}
& $\bar{\kappa} < 1.2\times10^{-3}$ \\
\hline

\makecell[l]{Finite $T_2$ \\ (Markovian spin depolarisation)}
& Spin $X/Z/Y$ errors
& 7
& \makecell[l]{0.36\% \\ {[Fig.~\ref{fig:all_err_stuff}b(i)]}}
& $T_2 > 417\tau_{\text{rep}}$ \\
\hline

\makecell[l]{Markovian ground-state dephasing \\ (Overhauser noise)}
& Spin $Z$ errors
& 10
& \makecell[l]{0.6\% \\ {[Fig.~\ref{fig:all_err_stuff}b(ii)]}}
& $T_2^* > 56\tau_{\text{round}}$ \\
\hline

Blinking
& Correlated loss
& 8
& \makecell[l]{6.1\% \\ {[Fig.~\ref{fig:all_err_stuff}b(v)]}}
& $P_A/P_D > 15.4$ \\
\hline

\end{tabular}

\caption{\textbf{Summary of fault-tolerance thresholds for realistic QD error sources in the sFFCC architecture.} For each physical error mechanism we give the equivalent error model at the resource-state generation circuit level, the maximum allowable number of physical fusion attempts $N$ at peak threshold, the corresponding threshold and required experimental benchmark. General loss considers all loss channels from state generation and detection, denoted by the end-to-end efficiency $\eta$. $C$ is the optical cyclicity, or minimum number of correct excitation-emission events required. $V^{ee}_{\text{HOM}}$ ($V^{se}_{\text{HOM}}$) refers to the emitter-emitter (single-emitter) Hong-Ou-Mandel visibility. Laser-induced spin-flip errors are represented by the normalised spin-flip rate $\bar{\kappa}$. The required $T_2$ is expressed in terms of photon generation time $2\tau_{\text{rep}}$ considering a Markovian spin depolarising noise channel. The required $T_2^*$ threshold is derived considering entanglement sources without an inherent spin echo. $P_A/P_D$ is the ratio of blinking rates between ``dead-to-alive" and ``alive-to-dead" events.}
\label{table:summ}
\end{table*}

\label{spin_depol}
We next consider finite spin coherence, $T_2$, originating from charge and nuclear noises. This was previously modelled as spin depolarising errors ($X$, $Y$, $Z$) after each gate in the Lindner-Rudolph protocol~\cite{Chan2025}. Here, we include its effect in the time-bin protocol, where each Pauli error is sampled from the Pauli group $\{I,X,Y,Z\}$ to occur after every step in the generation circuit with probabilities $\{1-p_{\rm dep},p_{\rm dep}/3,p_{\rm dep}/3,p_{\rm dep}/3\}$, i.e., after each optical excitation pulse, each spin $\pi$- and $\pi/2$-pulse, where $p_{\rm dep}$ is the spin depolarising error probability. 

Note that $X$-type errors that occur in the spin $\pi$-pulse \emph{between} time-bin excitations can lead to effective photon loss as well as spin errors. Since spin $X$ errors flip the state of the spin, photon emission is affected, resulting in either no emitted photon or double emission that equates to photon loss. In addition to possible effective loss, the location of an error event determines the exact nature of the remaining spin errors that propagate after the round; if a spin $X$ error occurs between the time bins in the case that a photon is emitted via the late excitation (a late time-bin photon), a spin $X$ error remains. If this happens in the case that a photon is emitted via the early excitation instead (an early time-bin photon), the spin not only experiences an $X$ error but also dephasing. If two spin $X$ errors occur between two time bins, i.e., one after the first excitation pulse and then again after the $\pi$-pulse, the $X$-error on the spin is reverted and photon loss is avoided, leaving only dephasing.

Fig.~\ref{fig:all_err_stuff}b(i) shows a set of thresholds that reaches a maximum of $p_{\rm dep}=0.36\%$ at $N=7$ and decreases thereafter, due to the same delocalised spin $X$ errors exhibited by branching errors. Assuming a Markovian noise model $p_{\rm dep}/3=\frac{1}{4}[1-\exp(-2\tau_{\rm rep}/T_2)]$~\cite{deGliniasty2024spinopticalquantum,Chan2025}, where $2\tau_{\rm rep}$ is the photon generation time in the time-bin protocol, this corresponds to a required minimum spin coherence time of $T_2=2\tau_{\rm rep}/(4\times0.36\%/3)=417\tau_{\rm rep}$. Given that $\tau_{\rm rep}=\tau_{\rm echo}=29$~ns (Table~\ref{tab:FOM}), $T_2>12.1~\mu$s.

Another noise source we consider originates from laser-induced spin-flip errors during spin rotation. The use of a red-detuned laser for driving spin rotation pulses in QDs has been shown to induce spin decoherence, as characterised by a linear increase in the spin-flip rate $\kappa$ with spin Rabi frequency $\Omega_R$, corresponding to a non-zero normalised spin flip rate: $\bar{\kappa}=\kappa/\Omega_r$~\cite{Meng2024,ChanThesis,appel_entangling_2022,Bodey2019}. For SiV centres, this is akin to heating-induced spin dephasing from microwave rotation pulses in resistive gold coplanar waveguides~\cite{Bhaskar2020}. 

Incoherent spin flips can appear during intra-bin rotation pulses as well as spin-echo refocusing pulses. For the former, if a spin-flip error appears during the spin $\pi$-pulse between early and late time-bin excitations, this results in the emission of zero (two) photons when the spin state is $\ket{0}$ ($\ket{1}$) during early time-bin excitation (Fig.~\ref{fig:all_err_stuff}a(iii)). In both cases, the resulting state is outside of the qubit space and equates to a lost photon in addition to $X$ and $Z$ errors on the spin (\SM C). This process has the same effect as having a spin $X$ error in the spin depolarisation model, which affects both photon loss and spin depolarising error thresholds. Therefore, we may bound our target $\bar{\kappa}$ for fault-tolerance with the stricter of the two thresholds, and set $\bar{\kappa} = 0.36\%/3$  since $\bar{\kappa}$ contributes to $1/3$ of the spin depolarising threshold. The required threshold is thus $\bar{\kappa} = 1.2\times10^{-3}$.

Next, we account for photon distinguishability. Successful type-II fusion gates rely on Hong-Ou-Mandel (HOM) interference~\cite{Hong1987}, requiring highly indistinguishable photons emitted from different or the same emitter. Here we describe the dominant noise processes that limit photon indistinguishability, and present their respective fault-tolerant thresholds. 

Emitter-emitter photon distinguishability predominantly originates from frequency mismatch between emitters. Slow fluctuations in the local charge or strain environment causes random drifts of the QD emission frequency with time, a phenomenon known as \emph{spectral diffusion}, typically occurring in millisecond timescales~\cite{Warburton2013} which is significantly longer than the QD lifetime ($235$~ps~\cite{Meng2024}). The effect of spectral wandering on emitter-emitter photon distinguishability has been examined previously and modelled as a Markovian photon $X$ error for $\{ZZ,XX\}$ fusions~\cite{Chan2025}, with a threshold of 4\% (Fig.~\ref{fig:all_err_stuff}b(vii)). 

So far, a threshold for single-emitter photon distinguishability has not been determined. Physically, interaction between the QD and nearby phonons in the semiconductor leads to incoherent photon emissions with broad phonon sidebands and broadening of the zero-phonon line~\cite{Tighineanu2018,PhysRevLett.93.237401}. While the phonon sidebands are removed by spectral filters which then manifest as photon loss, the latter contributes to pure dephasing of the QD excited state $\ket{2}$ due to elastic phonon scattering at a rate $\gamma_{d}$ (Fig.~\ref{fig:all_err_stuff}a(iii)). This degrades the indistinguishability $V=\Gamma/(\Gamma+2\gamma_d)$~\cite{Tighineanu2018, tiurev_fidelity_2021} of photons emitted from the same QD, where $\Gamma$ is the total decay rate. 

We model phonon-induced pure dephasing as stochastic $Z$ errors on the emitted time-bin photons (i.e., loss of early/late coherence). For rotated fusions that include Hadamard rotations, these $Z$ errors are mapped to $X$ in the fusion basis and therefore flip the equatorial-basis component of the fusion. In Fig.~\ref{fig:all_err_stuff}b(vi), we present the corresponding thresholds, which saturate at 0.57\% with $N=8$. The tolerable threshold for single-emitter distinguishability is lower than that for emitter-emitter distinguishability, since the former strictly impacts the encoded fusion outcome $\overline{XX} = (XX)^{\otimes n}$, which requires high-visibility interference between the early and late time bins emitted by the same QD over $n \leq N$ fusion attempts. In contrast, emitter-emitter distinguishability only affects the $\overline{ZZ} = ZZ$ outcome, which is recovered as soon as the first $ZZ$ outcome is obtained.


We now turn to the errors that may occur when we optically excite the target transition. During optical excitation of the cycling $|1\rangle\leftrightarrow|2\rangle$ transition with a $\pi$-pulse, two errors could occur depending on the excitation pulse duration~\cite{Appel2021_phdthesis, tiurev_fidelity_2021}: a short pulse may excite the off-resonant cycling transition $|0\rangle\leftrightarrow|3\rangle$ owing to partial spectral overlap, emitting a photon of unwanted frequency; a longer driving pulse may result in the emission of two photons in a time-bin due to re-excitation within the pulse. The pulse duration and shape may be optimised in light of this trade-off, and spectral filtering may be used to remove the dephased and distinguishable photons respectively. This dephasing reduces the indistinguishability, the effect of which has been analytically determined and illustrated by previous work~\cite{Chan2025}. Excitation error thresholds are thus captured by the distinguishability model in Fig.~\ref{fig:all_err_stuff}b(vii).

Photon loss, as well as errors, may also be correlated - here we describe the case where a charged QD may intermittently stop emitting photons (\textit{blinking}) due to slow tunnelling of carriers in and out of the QD~\cite{blinking,lodahl2015}, effectively leading to photon loss that is correlated in time. This results in bursts of sequentially lost photons in the photonic resource states. We model this non-Markovian blinking effect on the generation sequence by sampling each photon emission event based on the preceding event with probabilities for switching between non-emission to emission events, denoted as the ``dead-to-alive" $P_A$ and ``alive-to-dead" $P_D$ probabilities. Since each photon emission event depends on the state of the previous photon, the first photon in the sequence is sampled according to a stationary distribution $\{P_A,P_D\}$ for $P_A+P_D=1$. 

Threshold simulations in Fig.~\ref{fig:all_err_stuff}b(v) establish acceptable blinking rates, defining the corresponding loss rate for the ratio $P_A/P_D$ at which a threshold is achieved for a given $N$. Fig.~\ref{fig:all_err_stuff}b(iv) plots the uncorrelated photon loss threshold simulated in Ref.~\cite{Chan2025} which saturates at $8\%$ for $N=8$. This general bound encapsulates all sources of loss from photon generation to detection, and thus the effective loss tolerance for blinking depends on the remaining loss budget after considering, for example, outcoupling efficiency, fibre attenuation and insertion losses for optical elements, detector inefficiency, and finite $g^{(2)}$. 

For both loss models, we employ the re-initialisation technique introduced in Ref.~\cite{Chan2025}: any $ZZ$-only measurement projects the emitter spin into a known $Z$ eigenstate, effectively decoupling it from the resource state and re-initialising the spin, therefore the subsequent encoded RUS fusion can be re-attempted up to $N$ times. While this technique improves tolerance under uncorrelated loss, we find that it does not improve the blinking threshold. Intuitively, blinking induces correlated erasure bursts across multiple layers, such that no $ZZ$-only outcomes occur --- only fusion erasures. As a result, tolerance under blinking remains unaffected, saturating at $6.1\%$ for $N=8$, 24\% lower than the uncorrelated-loss threshold. This translates to a required blinking ratio $P_A/P_D>15.4$ for fault-tolerance.

Since blinking in charged QDs typically occurs on millisecond timescales, as evidenced in $g^{(2)}$ measurements at long time delays~\cite{ChanThesis}, and the resource-state generation completes within a few microseconds (Table~\ref{tab:qec_cycle_times}), the alive-to-dead probability $P_D$ is on the order of $10^{-3}$. On the other hand, $P_A$ is close to unity for a typical $n$-doped device which enables near-instantaneous QD charging with an electron ($\sim10$~ps)~\cite{Warburton2013}. In light of this, blinking is an insignificant effect for our particular system, though it may be relevant for other emitter-based platforms and hence must be included within the overall loss budget. 

Finally, we examine the effects of ground-state dephasing on our system. The fluctuating nuclear spin environment in the QD vicinity induces decoherence of the  spin qubit via the Overhauser effect~\cite{Glasenapp2016, Appel2021_phdthesis}, limiting the intrinsic spin dephasing time $T_2^*$. Here, we model its effects for emitter-based architectures which rely on resource-state generation protocols that do not have a built-in spin echo and are thus limited by $T_2^*$ (see discussion on comparing polarisation-encoded EPS with time-bin variant under Sec.~\ref{exp_capability_targets}). 

Previous work considers general dephasing errors of the spin qubit by applying a $Z$ error after each gate with probability $p_Z$ in the Lindner-Rudolph protocol~\cite{Chan2025}. This approach assumes a Markovian error model, whereas in reality, nuclear-spin noise is non-Markovian. We benchmark the Markovian approximation against a non-Markovian model, and verify that the non-Markovian model can be mapped to an effective Markovian model in the low-error limit ($p_Z\approx1\%$). Specifically, we find the fidelities of the Markovian model by simulating the resource-state generation circuit in Qiskit~\cite{qiskit2024}, and compare these to the analytical fidelities for the state's evolution under a normalised Gaussian magnetic field (see Methods).

\begin{figure}[hbtp!]
  \centering
  \includegraphics[width=0.9\linewidth]{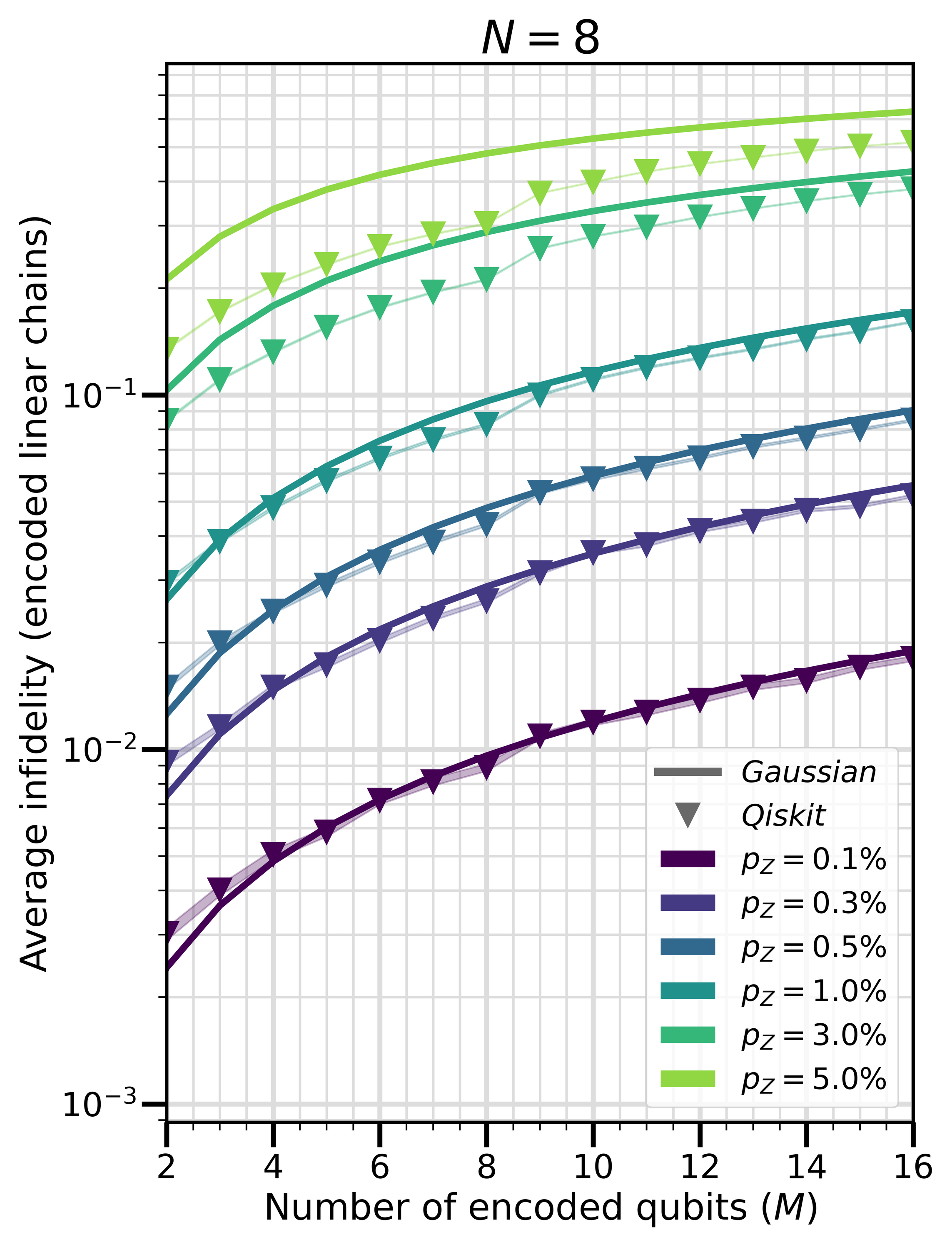}
  \caption{
  \textbf{Verifying Markovianity in the low-error limit for ground-state dephasing.}
    Average infidelity as a function of the number of encoded qubits $M$, for encoded linear cluster states with maximum allowable fusion attempt $N=8$. We observe good agreement between the simulated infidelity using Qiskit (triangle markers) and the analytical Gaussian envelope model (solid lines) in the limit of low error probability $p_Z<1\%$. Shaded region represents standard deviation of the simulated data.
  }
  \label{fig:qiskit_markov_sims}
\end{figure}

Fig.~\ref{fig:qiskit_markov_sims} shows excellent agreement between the two models when $p_Z<1\%$, with deviation at higher error rates, especially visible at $p_Z=3\%$. As the stochastic spin-$Z$ error approximation indeed suffices, we may refer to previous results~\cite{Chan2025} based on the polarisation-encoded Lindner-Rudolph protocol for a threshold that increases with the number of allowable physical fusions, saturating at 0.6\% after $N=10$ attempts (Fig.~\ref{fig:all_err_stuff}b(ii)).

For informing experiments, we translate each point in Fig.~\ref{fig:all_err_stuff}b(ii) into the minimum spin‑dephasing time \(T_2^*\) required to meet our error threshold. Using the relation $T_2^* =\frac{\sqrt{2}}{\Delta_{\rm OH}}$~\cite{Liu2007} and Eq.~(\ref{formula:delta_p}), we can express $T_2^*$ in terms of $\tau_{\rm round}$. For the peak threshold at $N=10$, this corresponds to $T^*_2=56\tau_{\text{round}}$. Initially, adding more photons per encoded qubit increases the protection against dephasing and lowers the required $T_2^*$, a benefit that is eventually outweighed by the increased wait time associated with more physical fusion attempts.

\subsection{Scaling and error budgets}
\label{scaling_budgets}

\begin{figure*}[hbtp!]
  \centering
  \includegraphics[width=\linewidth]{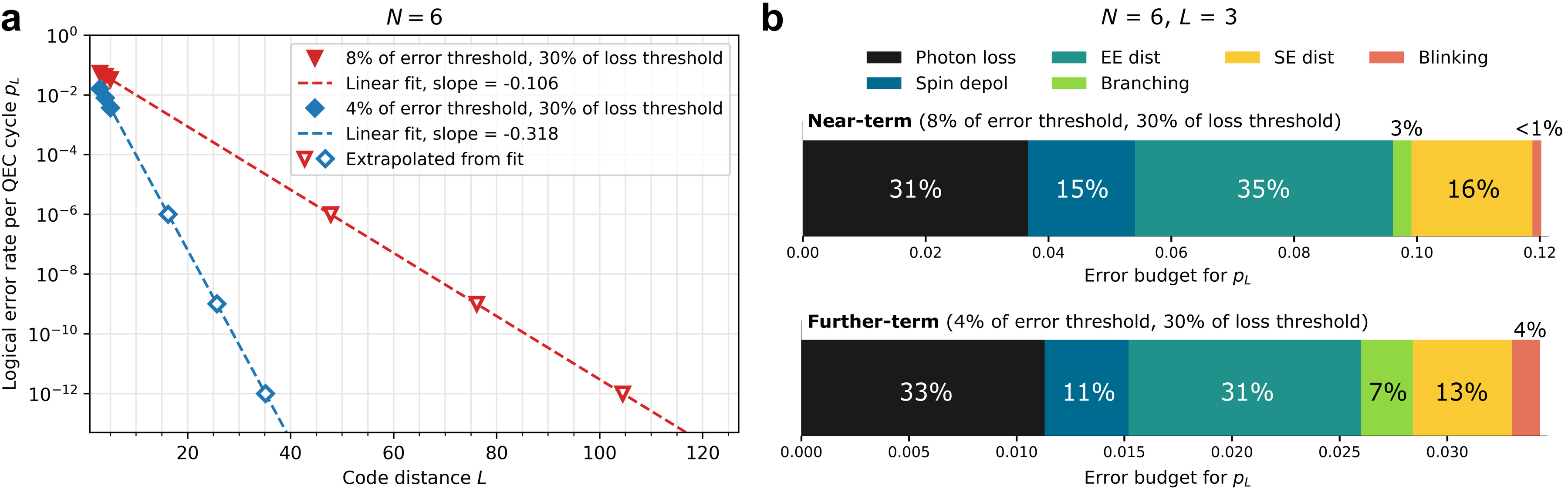}
  \caption{\textbf{Combined-noise scaling with all error sources active.} (a) Logical error rate per QEC cycle $p_L$ as a function of code distance $L$ at $N=6$ for two operating points: a near-term target ($8\%$ of each error threshold, $30\%$ of loss threshold) and a further-term target ($4\%$ of each error threshold, $30\%$ of loss threshold). In both cases, we set $P_D=10^{-4}$ and $P_A=1$, a physically motivated choice reflecting the negligible blinking typically observed in charged QDs. Triangle and diamond markers are simulated data. Dashed lines are least-squares fits, which are used to extrapolate code distances required to reach $p_L$ targets of $10^{-6}$, $10^{-9}$, and $10^{-12}$. 
  (b) Error budget for both operating points at $L=3$, $N=6$. Each bar shows the fractional contribution $\Delta p_L^{(i)} / \sum_j \Delta p_L^{(j)}$ of each error source, where $\Delta p_L^{(i)} = p_L - p_L^{(\backslash i)}$ is the reduction in logical error rate when source $i$ is removed while all others remain active.} 
  \label{fig:scaling}
\end{figure*}

The thresholds in Table~\ref{table:summ} are determined independently for each error source. In practice, all noise channels act simultaneously, and their combined effect determines the achievable logical error rate. We now quantify this through two complementary analyses: the scaling of the logical error rate with code distance under combined noise, and the relative contribution of each source. The interplay between noise sources is non-trivial. A particular example is the interplay between spin depolarisation and branching: when a spin $X$ error occurs between the two time bins, it prevents the photon emitted in the late time bin that would otherwise arise from branching.

First, we simulate the logical error rate per QEC cycle (one full round of error correction on the sFFCC lattice with all 6$L$ layers of fusions, syndrome extraction, and decoding), $p_L$, as a function of code distance $L$ with all noise channels active simultaneously at two representative sub-threshold operating points (Fig.~\ref{fig:scaling}(a)), for $N=6$. The suppression slope $\partial \log_{10} p_L / \partial L$~\cite{Fowler2012} is extracted by fitting $\log_{10} p_L$ versus $L$, and extrapolated to estimate the code distance required to reach target logical error rates of $10^{-6}$, $10^{-9}$, and $10^{-12}$. For a nearer-term target with each error source at $8\%$ of its individual threshold and photon loss at $30\%$ of its threshold; so that  $\sum_i p_i/p_{\mathrm{th},i} = 0.62$), we find a suppression slope of $-0.106$, requiring $L \sim 48$ for $p_L = 10^{-6}$, $L \sim 76$ for $p_L = 10^{-9}$, and $L \sim 105$ for $p_L = 10^{-12}$. For a further-term target with $4\%$ of error thresholds and same losses; $\sum_i p_i/p_{\mathrm{th},i} = 0.46$, the slope steepens to $-0.318$, requiring only $L \sim 16$ for $p_L = 10^{-6}$, $L \sim 26$ for $p_L = 10^{-9}$, and $L \sim 35$ for $p_L = 10^{-12}$. The contrast between the two shows how the error suppression of the code increases as the operating point moves further below threshold, where halving the errors alone triples the suppression slope, reducing the required code distance by a factor of ${\sim}3$. We emphasise that our parameter choices here are completely unoptimised, and therefore expected to be greatly improved. For instance, decreasing the error rates to $4\%$ of their thresholds allows us to reach the same performance with up to $50\%$ of the loss threshold. Most importantly, the photonic qubits are generated on the fly and measured immediately, so that the emitter count is the relevant hardware cost and the system does not require long-lived stationary qubits beyond the quantum-dot spin.

To identify which error source contributes most to the logical error rate at sub-threshold values, we compute an error budget following the method of Ref.~\cite{Google_below_threshold} (Fig.~\ref{fig:scaling}(b)). We simulate all error sources simultaneously at a given sub-threshold operating point, then repeat with each source removed in turn. The contribution of source $i$ is defined as $\Delta p_L^{(i)} = p_L - p_L^{(\backslash i)}$, and each bar shows the fractional share $\Delta p_L^{(i)} / \sum_j \Delta p_L^{(j)}$, representing the proportion of logical errors attributed to that source. For both aforementioned operating points, emitter-emitter distinguishability is the dominant contributor, accounting for $35\%$ and $31\%$ of the error budget respectively. Photon loss ($15\% \to 11\%$), spin depolarisation ($15\% \to 11\%$), single-emitter distinguishability ($16\% \to 13\%$), and branching ($3\% \to 7\%$) make up the remainder, while blinking contributes minimally. The dominance of emitter-emitter distinguishability at lower error rates reflects its shallow suppression slope; although its individual threshold is higher than other sources, the code corrects it less efficiently per unit reduction in physical error rate. This identifies emitter-emitter distinguishability as the primary bottleneck for scaling to low logical error rates, and suggests that improving photon indistinguishability between emitters will yield the largest gains in logical performance at these operating points. Of course, hardware improvements will not and need not occur at equal rates across the noise sources --- these plots serve as an illustrative example. Additionally, the combined threshold is substantially smaller than the sum of individual thresholds, which directly demonstrates that meeting each threshold independently is necessary but not sufficient, and all error rates must be brought well below their respective thresholds simultaneously to achieve fault-tolerance.

It is evident that there exist several opportunities for intelligent decoding given the distinct behaviour of some error mechanisms. As shown in the previous section, the fault-tolerant thresholds increase with the maximum number of physical fusion attempts up to $N=6$. Beyond this, they either begin to plateau or, as with branching and spin depolarising errors (error sources that include spin $X$ errors), start to decrease. As such, an intelligent decoder that flags incidences of $N>6$ as suspicious, where we may either terminate further attempts or use the information to increase the accuracy of the decoder by modifying the error probability weights in the syndrome graph, is beneficial. Therefore, the thresholds simulated here constitute worst-case estimates, with clear avenues for improvement.

\subsection{Resource Analysis}
\label{resource_scaling}
This section quantifies how physical resource requirements scale with the code distance $L$. Understanding the resource scaling is essential for, e.g., estimating the precise resources needed to achieve a target logical error rate inferred from sub-threshold simulations. To assess the space cost, we calculate the number of physical components required to implement the entanglement sources as well as the fusion measurement circuit. For the time cost, we compute the total time needed to perform all fusion measurements within one logical clock cycle, and compare our estimates directly with experimental benchmarks from other quantum computing platforms.

In the sFFCC lattice depicted in Fig.~\ref{fig:overview}a, there are $L^2$ hexagonal cells, each containing 6 linear cluster states, thus $6L^2$ independent entangled-photon sources are required. Each hexagonal unit cell has $6$ layers distributed in time, and 3 encoded fusions per layer. There are $L^3$ unit cells for a code distance of $L$. Thus, in one logical clock cycle (comprised of layers of physical clock cycles), there are in total $18L^3$ encoded fusions. For the further-term target specifications, $L=16$ is required to reach a logical error rate of $10^{-6}$ (Fig.~\ref{fig:scaling}a). Therefore, the hardware footprint is $6L^2 = 1{,}536$ QD emitters per logical qubit. For a rough comparison, the superconducting surface code requires $2d^2 - 1$ physical transmon qubits ($d \equiv L$ as the code distance); Ref.~\cite{Google_below_threshold} extrapolates $d = 27$ ($1{,}457$ qubits) to reach $p_L = 10^{-6}$ on their Willow processor. 

A key feature of our architecture is that it is compatible with excitation-based feedback. As described in Sec.~\ref{sec:classical}, in the case of photon loss, the next physical fusion is biased in $ZZ$, which requires both the active phase shifters in the unitary and fusion gates to be modified. The phase shifter in the latter must be configured to set the VBS to fully reflect in order to perform single-qubit measurements instead of two-qubit fusions. However, due to the reconfigurability of the EPS, it is possible to implement feedforward on the \emph{excitation pulses} rather than dynamically reconfiguring the measurement choice of each fusion gate in the detection. 

More precisely, on the \textit{excitation path} of one of the EPS involved in a fusion, we add an active optical element (e.g., fibre-coupled electro-optic modulator), which, through a short delay loop, selectively delays one pair of time-bin excitation pulses  before the biased fusion begins (Fig.~\ref{fig:timing}): When two time-bin photons do not overlap in time, they do not interfere at a beamsplitter and constitute deterministic single-qubit measurements. Note that to obtain the $ZZ$ outcome, the unitary gates of both EPS units still need to be adaptively switched to perform identity gates.

The key advantage of feedforward on excitation pulses is that all fusion gates can be \textit{fully passive}. Only one active element is needed at the excitation path of one of the EPS units, and the optical loss from this active element only lowers the available excitation power instead of the photon detection efficiency. In the time-bin implementation, this trick removes two loss channels (one 50:50 beamsplitter and one active phase shifter) per fusion gate. It is also well-known that on-chip passive optics have higher fidelity and lower loss than their active counterparts~\cite{optics_passive_active}. 

The delayed pulses must satisfy the inequality Eq.~(\ref{eq:ineq_EBF}). As shown in Fig.~\ref{fig:timing}, the delayed early time bin must arrive before the next spin $\pi$-pulse and is well resolved in time, thus $\tau_{\text{TB}}<\tau_{\text{EBF}}<\tau_{\text{int}}-\tau_{\pi}$, where $\tau_{\text{EBF}}$ is the pulse delay for excitation-based feedback, and $\tau_{\text{TB}}$ is the duration of the time-bin pulse dominated by the emitter's lifetime. 

We estimate the number of individual components of the fusion measurement circuit by counting the number of fusion gates (corresponding to edges of the lattice in Fig.~\ref{fig:overview}a), photon switches (vertices), and unitary gates (vertices), whereas the numbers of beamsplitters and phase shifters per component are extracted from Fig.~\ref{fig:component}a assuming time-bin encoding. In Table~\ref{table:resource}, we summarise the scaling of each physical element needed to prepare one logical qubit of code distance $L$. With the total number of fusion gates given by \(9L^2 -4L + 1\) (excluding boundaries), the number of various optical elements required for excitation-based feedback is reduced accordingly. In addition, \(3L^2\) fibre EOMs are added to the EPS units. Other smaller optics, such as mirrors, lenses and waveplates, are not considered.

\begin{table}[h]
  \begin{center}
    \begin{tabular}{|c|c|}
      \hline
      \textbf{Elements} & \textbf{Scaling with code distance $L$}  \\ 
      
        \hline
        \hline
        EPS units & $6L^2$ \\
        \hline
        Photodetectors & $18L^2$ \\
        \hline
        \multirow{2}{*}{Active phase shifters} & $33L^2-4L+1$; \\
         & $24L^2$ (with EBF)  \\
        \hline
        \multirow{2}{*}{Passive beamsplitters} & $54 L^2-8L+2$; \\
        & $45 L^2-4L+1$ (with EBF)\\
         \hline
                          \hline
        \textbf{Elements} & \textbf{Optical depth for a photon}\\
        \hline
        \hline

        Active phase shifters & $5$; $4$ (with EBF) \\
        \hline
        Passive beamsplitters & $6-8$; $5-7$ (with EBF) \\
                          \hline         
      
    \end{tabular}
    \end{center} 
 \caption{\textbf{Number of physical elements required to implement a logical qubit of code distance $L$}. Temporal fusions are assumed. Optical depth for a photon refers to the total number of respective optical elements a photon traverses before being measured~\cite{Clements2016}. EBF represents excitation-based feedback.}
    \label{table:resource}
\end{table}

\begin{figure}[h]
    \centering
	\includegraphics[width=1\linewidth, trim=0.cm 0cm 0.cm 0.0cm,clip]{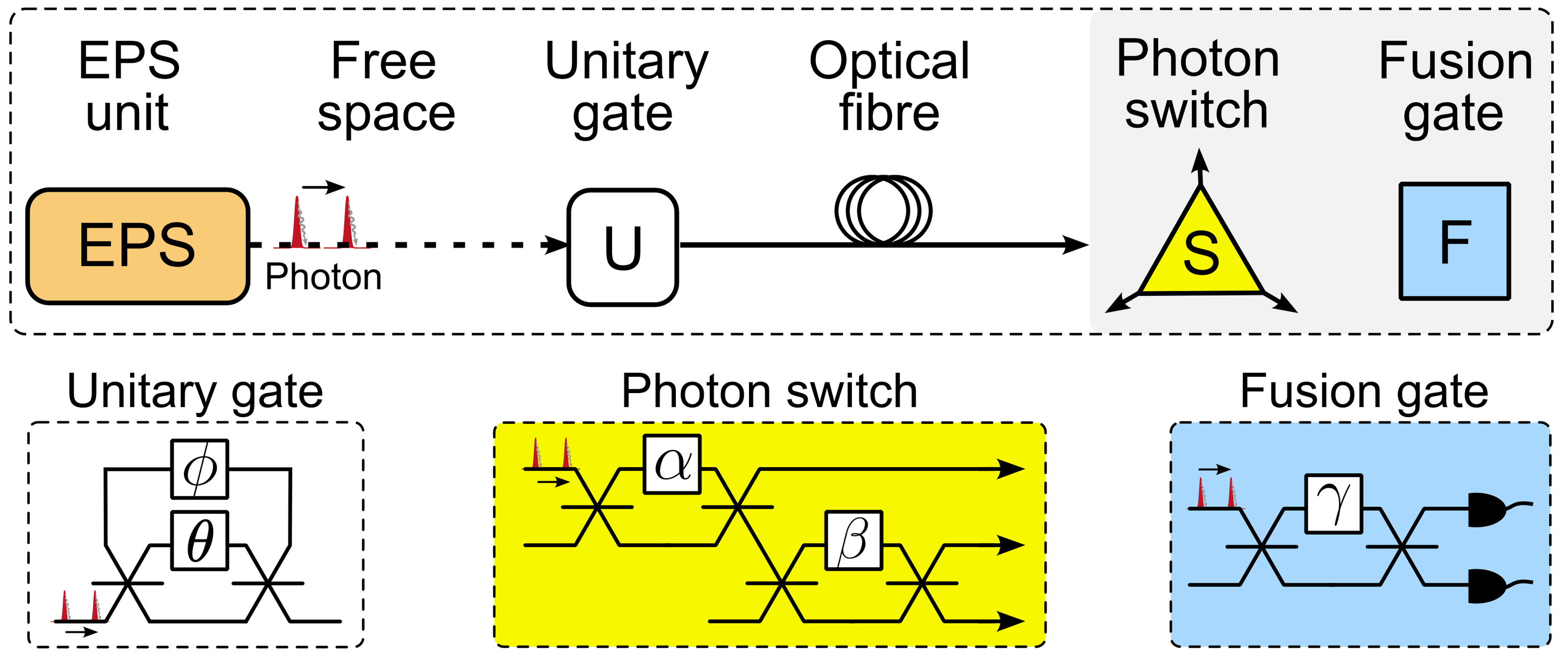}
    \caption{\textbf{Optical depth of the sFFCC architecture.} The EPS unit emits and outcouples a photon into a free-space unitary gate composed of two passive beamsplitters and two phase shifters. The photon is then coupled into an optical fibre, which directs it via a fibre array into a photon switch followed by a fusion gate, before arriving at a detector. Both the photon switch and fusion gate are fabricated on the same physical chip (grey region). Using excitation-based feedback, the fusion gate only requires a passive beamsplitter and two detectors (Fig.~\ref{fig:type}).}
    \label{fig:OD}
\end{figure}

Fig.~\ref{fig:OD} shows the optical depth per photon, defined by the total number of respective optical elements a photon traverses through from end to end~\cite{Clements2016}. Each photon emitted from an EPS unit travels through a free-space unitary gate, then couples into a single optical fibre, followed by a three-way photon switch, and finally a fusion gate, before reaching the detector. In terms of active elements (active phase shifters), the optical depth for each photon is 5, or $4$ if excitation-based feedback (EBF) is implemented. For passive elements, each photon passes through \(6\)–\(8\) beamsplitters depending on the routing path. With EBF, only one beamsplitter per fusion gate is used (Fig.~\ref{fig:type}), hence the optical depth is \(5\)–\(7\).

Comparing photonic architectures solely by the average number of photons per resource state or their loss thresholds can be misleading. A recent survey of existing FBQC proposals~\cite{psiq_loss_comparison} suggests that the average number of photons per resource state is a limited footprint metric, since it depends heavily on how resource states are prepared. 

In the sFFCC architecture, the total number of photons per resource state grows linearly with the lattice size; however, no additional optical components are required to generate these resource states -- only additional control pulses. Crucially, comparing photon number per state between sFFCC and multiplexed schemes with extensive switching network~\cite{psiq_loss_comparison} can obscure a key distinction: a higher loss tolerance achieved with the latter spreads the loss budget over \textit{many} additional channels, whereas the 8\% sFFCC loss threshold is governed by only a small number of optical components (Table~\ref{table:resource}). 

For instance, a recent work~\cite{Wein2024} utilises multiple emitters and RUS CZ gates to generate a Shor-encoded $(2,2)$ 6-ring state near-deterministically, which is then supplied to the 6-ring fusion network in Ref.~\cite{bartolucci_fusion-based_2023}. In this hybrid scheme, RUS acts as a form of temporal multiplexing during resource-state generation, with a loss threshold of 7.5\% applied over the optical depth across both state generation and the switch network. A major distinction of the sFFCC lattice is that its resource states are generated deterministically and are directly integrated into the fault-tolerant fusion network. Although the sFFCC lattice also employs temporal multiplexing in the form of RUS fusions, this occurs only at the point of measurement, forming an ``all-in-one" architecture with no intermediate RUS gates. As a result, it achieves optimal resource efficiency: one resource state is produced using only a single emitter per logical clock cycle. Moreover, the use of additional RUS gates and emitters in Ref.~\cite{Wein2024} is likely to introduce more errors during state generation, thus the practical error thresholds may be low.

Therefore, when drawing comparisons between architectures, it is important to contextualise the number of photons per resource state with the physical overhead, and the resulting distribution of error/loss channels across all components in the fusion network.

To estimate the time required to perform all fusion measurements within a single logical clock cycle, we carry out Monte Carlo simulations, described in Sec.~\ref{sec:errormodels}, where encoded fusion outcomes are simulated layer by layer. The physical clock cycle time for a given layer is obtained by simulating the maximum number of physical fusion attempts required by emitters on the layer, while the logical clock cycle time is the sum of the physical clock cycle times across all layers. Photon loss is the only error channel that directly impacts the physical clock cycle due to biased fusion. In the presence of loss, biasing the next attempt in $ZZ$ reduces the time needed to complete an encoded fusion, since only one successful encoded fusion outcome is required which can be deterministically recovered through single-qubit measurements. As discussed in Sec.~\ref{sec:classical}, emitters that complete their encoded fusions early are idled via buffer spin $\pi$-pulses until all emitters are ready to proceed to the next layer. 

\begin{figure}[h]
	\includegraphics[width=1\linewidth, trim=0.cm 0cm 0.cm 0.0cm,clip]{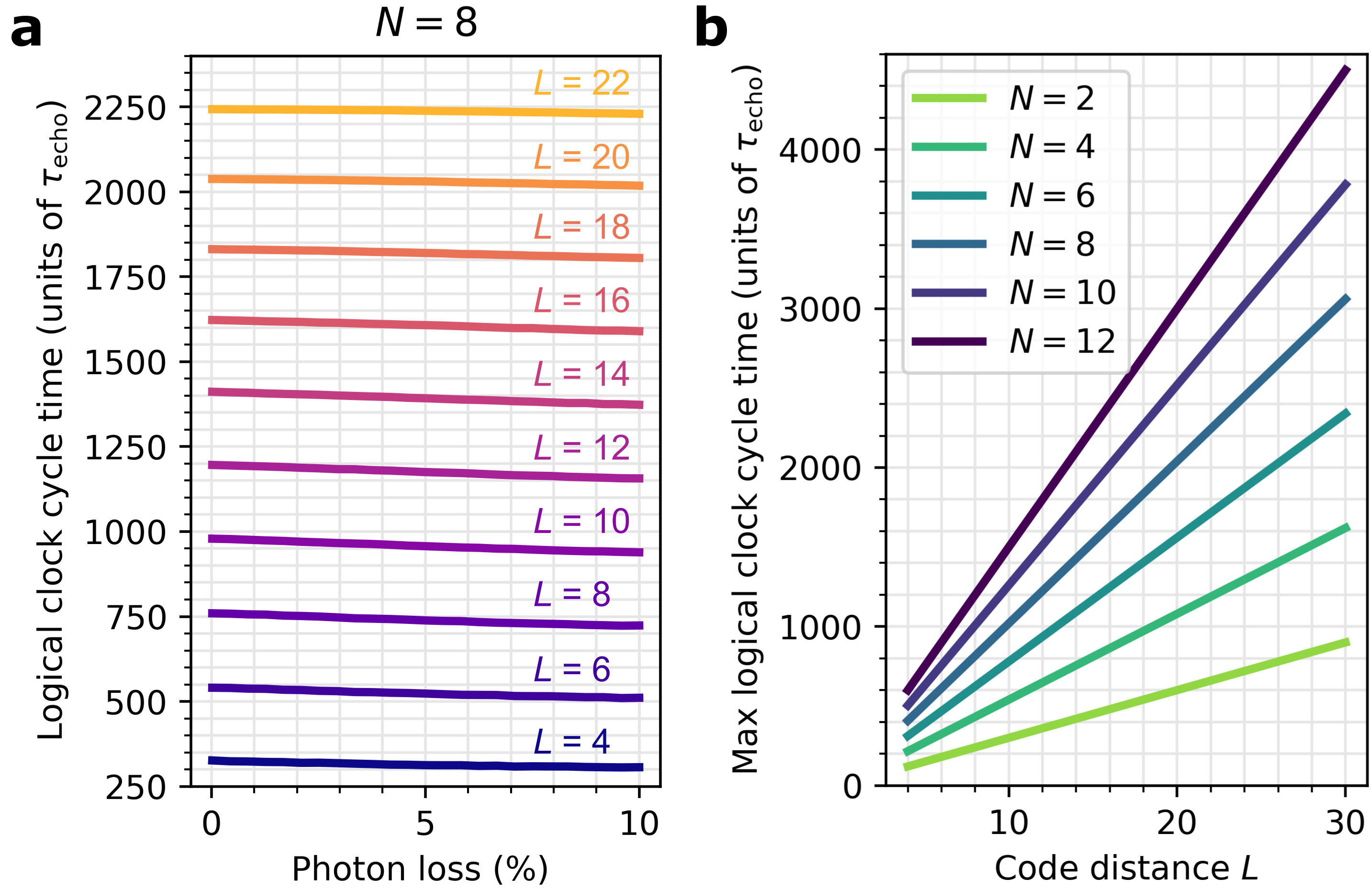}
    \centering
    \caption{\textbf{Logical clock cycle time $\tau_{\rm logical}$ in units of $\tau_{\text{echo}}$.} \textbf{(a)} Simulated $\tau_{\rm logical}$ as a function of photon loss and lattice size $L$. $\tau_{\rm logical}$ decreases with higher loss due to the use of biased fusions which reduces the physical clock cycle time.
    \textbf{(b)} Plot of Eq.~(\ref{eq:logicaltime}) with $n^i_{\rm max}=N$.}
    \label{fig:time}
\end{figure}

Therefore, the duration of physical clock cycle on layer $i$ is given by $(2n^i_{\rm max}+1)\tau_{\rm echo}$, where $n^i_{\rm max}\leq N$ is the maximum number of physical fusion attempts among all emitters on the layer, and each photon takes $2\tau_{\rm rep}=2\tau_{\rm echo}$ (Eq.~(\ref{eq:ineq1})) to generate. Another $\tau_{\rm echo}$ is required for applying a spin $\pi/2$-pulse to terminate the encoded fusion (Fig.~\ref{fig:control}). The logical clock cycle time is then
\begin{align}
    \tau_{\rm logical} = \sum^{6L}_{i=1}(2n^i_{\rm max}+1)\tau_{\rm echo}.\label{eq:logicaltime}
\end{align}
Fig.~\ref{fig:time}a presents the simulated logical clock cycle time $\tau_{\rm logical}$ as a function of photon loss and lattice size $L$ at $N=8$. As seen in the figure, the logical clock time $\tau_{\rm logical}$ increases with the lattice size $L$ as we also consider a larger lattice in the temporal dimension. Furthermore, as the loss rate increases, we see a slight reduction in $\tau_{\rm logical}$ since a photon loss is followed by a biased fusion, which has a higher success probability. At lower values of $L$, the reduction in $\tau_{\rm logical}$ due to biased fusion is more prominent. It is less significant at a higher $L$ since the probability for at least one encoded fusion on any layer $j$ to end up with only $\overline{XX}$ outcome (thus $n^j_{\rm max}=N$) increases with more emitters. In general, we observe an almost linear dependence of $\tau_{\rm logical}$ on $L$. Fig.~\ref{fig:time}b plots the absolute maximum $\tau_{\rm logical}$ where $n^i_{\rm max}=N$ as a function of $N$ and $L$.

\begin{table*}[hbtp!]
  \centering
  \resizebox{\textwidth}{!}{%
    \begin{tabular}{|c|c|c|c|}
      \hline
      \textbf{Platform} & \textbf{QEC code (distance)} & \textbf{Logical cycle time} & \textbf{Exp./theory (year)} \\
      \hline\hline
      Google (superconducting) & Surface code ($d = 5, 7$) & 1.1~$\mu$s & Experimental (2024)~\cite{Google_below_threshold} \\
      \hline
      IBM (superconducting) & Heavy-hexagon code ($d = 3$) & 1.5~$\mu$s & Experimental (2023)~\cite{ibm_heavyhex} \\
      \hline
      Harvard/QuEra (neutral atoms) & 7-qubit Steane code ($d = 3$) & 0.5~ms & Experimental (2023)~\cite{atom_arrays} \\
      \hline
      Quantinuum (trapped ions) & 7-qubit Steane code ($d = 3$) & 200~ms & Experimental (2021)~\cite{quantinuum_honeywell} \\
      \hline
      This work (QD-based photonics) & sFFCC ($L = 3$) & $6~\mu$s & Theoretical \\
      \hline
    \end{tabular}%
  }
  \caption{\textbf{Logical clock cycle times across different physical platforms.} One logical clock cycle includes the duration of one round of syndrome extraction, made up of layers of physical quantum error correction (QEC) cycles, and excludes classical decoding. For InAs QDs, the optimal choice of $\tau_{\text{echo}}$ is 29~ns (Table~\ref{tab:FOM}), bringing the logical clock cycle time to a total of $209\tau_{\text{echo}}\approx6~\mu s$ for $L=3, N=8$ and 8\% photon loss (Fig.~\ref{fig:time}).
  }
  \label{tab:qec_cycle_times}
\end{table*}

For a wider comparison, we benchmark our estimation of the logical clock cycle time with other quantum computing platforms, as shown in Table~\ref{tab:qec_cycle_times}. Assuming that the optimal spin-echo delay $\tau_{\rm echo}\approx29$~ns~\cite{Meng2024}, $\tau_{\rm logical}\approx6~\mu$s for $L=3$, $N=8$ and a photon loss of $8\%$. Our blueprint projection achieves a competitive, microsecond-fast logical clock cycle speed which scales linearly with $L$. 

Note that for GaAs QDs in an AlGaAs matrix grown with local droplet etching~\cite{Nguyen2023}, $\tau_{\rm echo}$ can be chosen to be smaller for a faster speed, owing to a slow decay of the spin-echo visibility with $\tau_{\rm echo}$, but the relevant timing requirements (Eqs.~(\ref{eq:ineq1})-(\ref{eq:ineq5})) still need to be fulfilled. Moreover, the addition of ancilla-assisted RUS fusions (see further discussion in Sec.~\ref{exp_capability_targets}) would impose a trade-off between the logical clock cycle time and photon loss tolerance: A high loss can be tolerated by re-attempts of encoded fusions, but at the cost of slowing down the logical clock cycle.

\subsection{Logical operations for computation} 
\label{open_q}
Naturally, the next step in designing an architecture would be to carry out logical operations for computation. Methods for performing different gates on topological codes include the use of code deformation and manipulations of twist defects to realise the full set of Clifford gates~\cite{pokingholes, logicalblocks}. These methods can be directly adapted for Floquet codes, as the instantaneous stabiliser group of each round is merely that of the toric code. Another technique achieves the same outcome by selecting appropriate measurement sequences to create spacetime symmetry defects, enacting the logical gates over the natural evolution of the code~\cite{davydova2024}. Additionally, universal quantum computation must include non-Clifford gates~\cite{gottesman_knill}. This is usually approached via the injection of noisy so-called \textit{magic states} into the code, that must then be distilled~\cite{magicstatedistillation, Gidney2019efficientmagicstate, Litinski2019magicstate, hirano2024} or cultivated~\cite{magicstatecultivation} until the output magic state reaches sufficient fidelity. 

The non-Clifford state typically used is $\ket{T} = Z^{\tfrac{1}{4}} \ket{+}$ as it can be injected at relatively high fidelity~\cite{gidneyhook}, distilled via increasingly compact circuits, used four-at-a-time to implement a Toffoli gate~\cite{cody2013} (essential for any complex classical computations), and may be corrected by cheap Clifford operations after teleportation. An alternative is the $\ket{CCZ} = CCZ \ket{+++}$ state; although more difficult to prepare, only one is needed for a Toffoli gate. Recent work has examined using domain walls between the twisted quantum double model and toric codes to achieve logical \textit{T} and \textit{CCZ} gates in 2D~\cite{quantumdoublemodel}. Paired with a just-in-time decoder, the gates are fault-tolerant, and avoid the usual distillation/cultivation route. The framework outlined by the authors in that work may be readily applied to our sFFCC architecture, with its aforementioned toric code slices in time. We note also a similar approach that achieves a different non-Clifford gate in~\cite{sajith2025}. The thresholds obtained herein for our architecture (Sec.~\ref{sec:errormodels}) do not account for the effects of such topological features, nor have we simulated non-periodic boundaries~\cite{fowler2013}; this will form the focus of future work.

\subsection{Experimental capabilities and thresholds}
\label{exp_capability_targets}
So far, we have presented the simulated fault-tolerant thresholds for error sources present in the experimental platform. Benchmarking these thresholds against current state-of-the-art experimental capabilities identifies key areas where engineering efforts should be consolidated for scalable quantum error correction. 

For photon loss, the relevant loss channels over the optical path of a photon can be categorised into three groups: (i) losses within the EPS unit, from photon emission into the broad phonon sideband ($0.223$~dB~\cite{Uppu2020}) to coupling loss into an optical fibre ($3$~dB~\cite{Loredo2026}); (ii) interface losses between the EPS unit and the fusion measurement circuit chip, which includes fibre transmission loss and fibre-to-chip coupling from a fibre array ($0.13$~dB)~\cite{Alexander2024}; and (iii) losses within the fusion measurement circuit itself, including insertion losses from each active phase shifter ($0.53~\rm{dB~cm^{-1}}$) and passive beamsplitter, waveguide propagation losses ($2~\rm{dB~m^{-1}}$), and on-chip photon-number resolving detector inefficiency ($0.3$~dB)~\cite{Alexander2024}. For a photon loss threshold of $8\%$ (Table~\ref{table:summ}), the total loss must be less than $\sim0.362$~dB.

Fault tolerance requires an optical cyclicity of $C>574$. In the Voigt magnetic field where optically driven spin transitions are accessible, the current highest measured value on the QD platform is $C\approx36$~\cite{Meng2024}. This might be improved, i.e., by tuning the external magnetic field with low-strain QDs. Alternatively, $C\approx600\pm200$~\cite{Antoniadis2023} has been measured in a Faraday magnetic field, suggesting that a combination of high cyclicity enabled by the Faraday geometry and spin control through microwave pulses by a planar antenna could be a viable route.

All other error sources in Table~\ref{table:summ} have been shown to have error rates either close to or below thresholds: raw emitter-emitter HOM visibility $V^{ee}_{\text{HOM}}\approx93\%$~\cite{Zhai2022} and raw single-emitter HOM visibility $V^{se}_{\text{HOM}}\approx97\%$~\cite{Loredo2026} have been demonstrated. The optically induced spin-flip rate has been observed to be as low as $\bar{\kappa}\approx3\times10^{-3}$~\cite{Meng2024}. A spin coherence time $T_2\approx113~\mu$s~\cite{Zaporski2023} has been achieved with dynamical decoupling, which is inherent to the generation of encoded linear cluster states (Fig.~\ref{fig:rsg}b). Blinking for charged QDs is negligible.

Now, we further motivate the choice of time-bin encoding for EPS units over the polarisation-encoded variant. In the latter approach~\cite{Cogan2023,Huet2025,laccotripes2025entangledphotonsourcetelecom}, a fundamental trade-off exists between single-emitter photon indistinguishability and coherent spin control. These sources typically rely on a four-level system with two degenerate optical transitions~\cite{lindner_proposal_2009}, where spin control is achieved by applying a small magnetic field (on the order of a few mT) to induce Larmor precession of the spin qubit. However, due to the Zeeman effect, the two vertical transitions acquire slightly different optical frequencies, reducing the spectral overlap between the circularly polarised $\sigma_+$ and $\sigma_-$ photons. This spectral mismatch degrades photon indistinguishability and lowers fidelity of the resource states. In contrast, time-bin EPS units eliminate this trade-off, as the same optically cycling transition is used for both early and late time-bin emissions~\cite{Meng2024}.

Polarisation-encoded EPS units based on QDs are typically not equipped with spin echo. Without spin echo, the time to emit an encoded qubit of $N$ photons needs to be shorter than the emitter's spin dephasing time $T^*_2$ (see considerations of ground-state dephasing in Sec.~\ref{sec:errormodels}). Therefore, to implement feedforward for RUS fusions, this demands fast modulators and electronics with latency far below $T^*_2/N$ for optical switching between consecutively emitted photons, given that the electron (hole) spin dephasing time has been observed to be $T^*_2\approx2$~ns (20~ns)~\cite{Huet2025}.
 
While ultrafast optical switching has been demonstrated~\cite{fenwick2025ultrafastswitchingtelecomphotonnumber,Kupchak_2019}, it is unclear whether electronic feedback on this timescale is possible due to jittering. Furthermore, the propagation-delay from EPS to detectors also needs to be included in the overall feedforward time budget, which is difficult to meet without monolithic integration given that photons travel around 30 cm per nanosecond. With spin echo, these problems are partially mitigated. Doing so would add extra refocusing pulses and latency, extending the feedback budget. Nevertheless, sub-nanosecond switching times with a switching resolution~\cite{fenwick2025ultrafastswitchingtelecomphotonnumber} larger than the photon pulse width might still be necessary. While $T_2^*$ may be prolonged by nuclear cooling~\cite{Huet2025,laccotripes2025entangledphotonsourcetelecom}, it is unlikely that this would work at low magnetic fields in which nuclear Overhauser broadening dominates.

Two strategies that have been shown to reduce the resource cost of photonic fault-tolerant quantum computation are \textit{active volume} compilation~\cite{active_volume} and \textit{interleaving}~\cite{bombin_interleaving_2021}. Below we briefly discuss the compatibility of our blueprint with each.

The active volume architecture~\cite{active_volume} requires non-local connections between logical qubit \emph{modules}, i.e. addressable logical locations that can store one logical qubit in a surface code patch. Given \(N_Q\) total qubit modules, each one must be connected to \(O(\log N_Q)\) other modules to enable quickswap operations and lattice surgery between non-neighbouring patches. Our sFFCC architecture is compatible with this approach. Since each logical qubit in our scheme is encoded across an array of EPS units via a planar fusion measurement circuit with reconfigurable photon switches, long-range connectivity between logical qubits can be established by routing photons from one logical qubit's EPS units to fusion gates associated with another logical qubit. This enables parallelisation of logical operations, such as those required for algorithms like Shor's factoring, where the active volume can be orders of magnitude lower than the circuit volume. Therefore, logical qubits can undergo quickswap-like operations and lattice surgery within a finite range through the reconfigurability of the photon switches and routing network already present in our blueprint. A detailed analysis of the active volume cost for specific algorithms in the context of our architecture is thus an interesting direction to explore.

Interleaving~\cite{bombin_interleaving_2021} reduces the number of RSGs by time-multiplexing a single RSG with photonic delay lines to generate multiple logical qubits sequentially. In our architecture, the EPS units play the role of RSGs. However, a direct implementation is not straightforwardly compatible with the adaptive RUS fusion strategy used in our scheme. RUS fusion requires real-time classical feedback from fusion outcomes to the EPS pulse sequences, determining whether to repeat an attempt, bias the next attempt, or move on to the next encoded fusion. With photons stored in delay lines for interleaving, this feedback loop is broken, since the EPS unit must already commit to the next emission before the previous fusion outcome is known. Interleaving can nevertheless be made compatible with emitter-based architectures by truncating the one-dimensional cluster-state chains to finite length~\cite{paesani_high-threshold_2022}. In this approach, each EPS unit generates a finite resource state, such as a short chain or branched chain of length $l$, which is then stored in a photonic delay line and later fused in the measurement circuit. The fault-tolerance threshold degrades only modestly with chain length; even moderate values such as $l=14$ recover a good threshold relative to the infinite-chain limit. With finite chains, the architecture can no longer use adaptive RUS fusion, but it can still employ static repetition-code encoded fusions as in previous work~\cite{Chan2025}, which continue to protect against fusion failure and photon loss, albeit with lower thresholds than the adaptive RUS scheme. This creates a resource--threshold trade-off: interleaving reduces the number of physical EPS units, and therefore the hardware footprint, but at the cost of a lower fault-tolerance threshold. Determining the best operating point of this trade-off for a target logical error rate is an important direction for future work.

Finally, we highlight a modification to our architecture that would offer a unique advantage over optical fusion-based architectures~\cite{psiq_loss_comparison} --- that of ancilla-assisted encoded RUS fusion. Ancilla-assisted entanglement schemes have been extensively discussed~\cite{Nemoto2014, distributed_silicon,Choi2019, Sharman2021, Lukin_nRUS, Kim2014, Dolde2014, Lim2005, Lim2006, Sharman2021quantumrepeaters, dyte2025storingquantumcoherencequantum, Bruschi2014}, where the fast, optically accessible electron spin of a colour centre or QD emitter is used to perform repeated entanglement attempts mediated by photons, before swapping the entanglement to the much slower but highly coherent nuclear spin memory. Using fusion gates for entanglement generation, if the fusion fails or either of the two photons is lost, the electron spins can simply be re-initialised and the fusion is retried indefinitely until a successful click pattern occurs. As soon as fusion succeeds, the resulting electron-photon entanglement is swapped onto the nuclear spin, leveraging its long coherence time. Since the electron‐spin-mediated fusions can be repeated until a successful outcome is obtained, this scheme is \emph{fully tolerant} to photon loss. 

The trade‐off is that (i) the overall clock rates are slowed while waiting for fusions to succeed, (ii) as time passes, the nuclear spin decoheres, and (iii) the effect of loss is effectively converted into errors, demanding very low spin noise. In practice, this requires the addition of an ancilla spin qubit for each emitter and high-fidelity electron-nuclear entanglement swapping.

For QD platforms, there are ongoing efforts to realise a nuclear ancilla qubit by polarising the nuclear spin ensemble~\cite{Appel2025}. It is, however, an outstanding feature that our architecture is inherently and exclusively compatible with ancilla‐assisted RUS fusions.

\section*{Discussion}
\label{conclusion}
We have proposed a realistic blueprint for constructing a fault-tolerant, low-depth, and modular photonic architecture with semiconductor quantum dots (QDs) that offers practical solutions to the main challenges facing photonic quantum computing. 

Deterministic entangled-photon sources (EPS), adaptive repeat-until-success (RUS) fusions, and the synchronous foliated Floquet colour code (sFFCC) fusion network synergistically integrate to achieve both scalability and resource efficiency: Time-bin encoded EPS coupled with temporal fusions and low connectivity of the lattice minimise the optical depth per photon and remove the need for spatial multiplexing. Excitation-based feedback at the EPS level enables fully passive fusion gates, reducing active optical elements and loss channels. 


With noise models that translate QD-specific error sources into fault-tolerance thresholds with clear experimental benchmarks, and by matching precise resource and timing requirements to current device performance, our blueprint bridges the gap between theoretical design and experimental implementation. While focused on semiconductor QDs, our framework is applicable to other emitter platforms~\cite{Chan2025} which have access to spin and Purcell-enhanced optical transitions, such as single neutral atoms trapped in an optical cavity~\cite{Thomas2022,Thomas2024,Grinkemeyer2025} and colour centres~\cite{colour_centres_review,Bhaskar2020,fischer2025spinphotoncorrelationspurcellenhanceddiamond}.

We build upon components that have already been experimentally demonstrated and thoroughly characterised, providing a clear roadmap toward the realisation of a functional logical qubit in photonic quantum computing. With the remaining areas for improvement now well defined, efforts may gradually begin to shift from primarily fundamental research toward a focus on targeted engineering development. 

\section*{Methods}
\subsection*{Threshold simulations}
We adopt the same decoding methods used previously~\cite{paesani_high-threshold_2022, Chan2025}. In the case of photon loss, the fusion outcomes are completely erased (and heralded as two-photon detection events are required). We then create \textit{supercells} by merging adjacent stabilisers with Gaussian elimination which results in new correlation surfaces that percolate between the logical operators while avoiding lost fusions~\cite{supercell}; if such a surface is not found, this results in logical erasure. We can decode error strings in the fusion syndrome lattice with PyMatching~\cite{higgott2023sparse}. A fusion error captures the effect of the Pauli errors on the involved photons, and a fusion erasure likewise captures the loss of the photons as well as the probabilistic failure of the gate. Our simulations use Monte-Carlo sampling where the relevant physical error rates are varied to estimate the error rates of a logical qubit for various lattice sizes or code distances $L$. For each set of error parameters, we perform $10^4$ trials of quantum error correction with the sFFCC architecture under periodic boundary conditions to produce thresholds for the bulk lattice, fixing the fusion failure rate to 50\% throughout. Each threshold is obtained for each error independently, setting the other error rates to zero. When simulating the target parameter space for fault-tolerance under three simultaneous sources of error, namely loss, branching error, and single-emitter distinguishability, we sample these in that order for each threshold obtained. 

The main result is summarised in Table~\ref{table:summ}, and a complete set of threshold curves obtained for each noise model as a function of the maximum number of fusion attempts $N$ is compiled in Fig.~\ref{fig:all_err_stuff}. 

\subsection*{Fidelity of encoded linear cluster state under ground-state dephasing} 
Consider an ideal encoded linear cluster state of $M$ encoded qubits, each encoded with $N$ photons, 
\begin{equation}
\label{eq:psi0}
\ket{\psi_0}
=\frac{1}{\sqrt{2^M}}
\sum_{x\in\{0,1\}^M}
(-1)^{f(x)}\,\ket{x}_E,
\end{equation}
where each basis state $\ket{x}_E$ of the encoded qubit is labelled by a bit-string $x = (x_1,\dots,x_M)\in\{0,1\}^M$. Each basis state's Hamming weight,
$\sigma(x)=\sum_{j=1}^M x_j$, counts the number of 1's (photons) in the string. The normalised cluster state is then the equal-weight superposition of all \(2^M\) such strings, each carrying a phase \((-1)^{f(x)}\), where $f(x)=\sum_{j=1}^{M-1}x_j\,x_{j+1}\bmod2$
counts the number of adjacent “11” pairs and thus describes the characteristic entanglement pattern of a linear cluster state. If $\ket{\psi_0}$ experiences noise from a static, uniform magnetic field with some angular frequency shift $B$, it can be rewritten as
\begin{align}
\label{eq:psiB}
    \ket{\psi(B)}
    =\sum_{x\in\{0,1\}^M}
    \frac{(-1)^{f(x)}}{\sqrt{2^M}}
    e^{-iNB\tau_{\mathrm{round}}\,\sigma(x)}\,
    \ket{x}_E,
\end{align}
where $\tau_{\text{round}}$ is the time taken to emit a polarisation-encoded photon. The density matrix is thus given by 
\begin{align}
\label{eq:rhoB}
\rho(B)
    &= \ket{\psi(B)}\bra{\psi(B)} \nonumber\\
    &= 
      \sum_{x,y\in\{0,1\}^M}
      \frac{(-1)^{f(x)-f(y)}}{2^M}\nonumber \\
    &\quad\quad\times
      e^{-iNB\tau_{\mathrm{round}}
        \bigl[\sigma(x)-\sigma(y)\bigr]}
      \ket{x}_E\bra{y}_E.
\end{align}
To account for the Overhauser effect from a nuclear spin bath, the magnetic field is randomly fluctuating; we therefore modify $\rho$ with a Gaussian distribution $p(B) = \mathcal{N}(0,\Delta_{\rm OH})$ for root-mean-square fluctuations $\Delta_{\rm OH}$, and integrate to obtain:
\begin{align}
\label{eq:rho}
\rho &= \int_{-\infty}^\infty p(B)\rho(B)dB \nonumber\\
&= 
  \sum_{x,y\in\{0,1\}^M}
  \frac{(-1)^{f(x)-f(y)}}{2^M}
  \nonumber\\
&\quad\quad \times e^{-\tfrac12(N\tau_{\mathrm{round}}\Delta_{\rm OH})^2
    \bigl[\sigma(x)-\sigma(y)\bigr]^2}\ket{x}_E\bra{y}_E.
\end{align}
From here we calculate the state fidelity $\mathcal{F}$, which reduces quadratically with both the state preparation time and the number of photons per encoded qubit:

\begin{equation}
\label{eq:fidelity}
\begin{aligned}
\mathcal{F}
&= \bra{\psi_0}\rho\ket{\psi_0} \\
&= \frac{1}{2^{2M}}
   \sum_{x,y\in\{0,1\}^M}
   e^{-\tfrac12(N\tau_{\mathrm{round}}\Delta_{\rm OH})^2
     \bigl[\sigma(x)-\sigma(y)\bigr]^2}.
\end{aligned}
\end{equation}

To directly compare the Gaussian model with the Qiskit simulations, $\Delta_{\rm OH}$ must be found in terms of $p_Z$ by considering the density matrix $\rho_Z(N)$ of the noisy state after $N$ rounds of photon emissions per encoded qubit:
\begin{equation}
\rho_Z(N) = \begin{pmatrix}
  1 & 1-2Np_Z\, \\[6pt]
  1-2Np_Z\, & 1
\label{eq:p_z}
\end{pmatrix},
\end{equation}
where $p_Z$ is the per-gate error applied after each photon emission (i.e., after a $C_{\text{NOT}}$ gate) and after every Hadamard gate in Qiskit. This is done by equating the off-diagonal elements of the respective density matrices Eq.~(\ref{eq:rho}) and $\rho_Z(N)$ to obtain: 
\begin{gather}
\Delta_{\rm OH}^2 = \frac{-2\ln(1 - 2Np_Z)}{\tau_{\mathrm{round}}^2 N^2},
\label{formula:delta_p}
\end{gather}
which gives us the equivalent $\Delta_{\rm OH}$ for an input $p_Z$ when calculating Eq.~(\ref{eq:fidelity}).

\section*{Author contributions}
The project was conceived by P.L.~~M.L.C and A.A.C. contributed equally to this work.~~M.L.C designed the blueprint with inputs from A.S.S and S.P.~~A.A.C performed error modelling and threshold simulations with inputs from A.S.S and S.P.~~S.P, A.S.S, and P.L supervised the project.~~M.L.C and A.A.C wrote the manuscript with inputs from all authors.

\section*{Acknowledgements} 
We thank Martin Hayhurst Appel and Matthias Löbl for fruitful discussions.
We gratefully acknowledge financial support from Danmarks Grundforskningsfond (DNRF 139, Hy-Q Center for Hybrid Quantum Networks), the Novo Nordisk Foundation (Challenge project ``Solid-Q”), and Danmarks Innovationsfond (IFD1003402609, FTQP). M.L.C acknowledges funding from Danmarks Innovationsfond (Grant No.~4298-00011B). A.A.C. acknowledges funding from UK EPSRC (EP/SO23607/1). S.P. acknowledges funding from the Marie Skłodowska-Curie Fellowship project QSun (Grant No. 101063763), the VILLUM FONDEN research grants No.~VIL50326 and No.~VIL60743, and support from the NNF Quantum Computing Programme. 

\section*{Competing Interests} 
P.L. is the founder of the company Sparrow Quantum which commercialises single-photon sources. M.L.C is an employee at Sparrow Quantum. The other authors declare no competing financial or non-financial interests.

\section*{Data Availability}
The dataset generated for this article is openly available at the GitHub repository~\cite{github}.

\section*{Code Availability}
The code that generates the threshold and resource estimation data of this article is openly available at the GitHub repository~\cite{github}.

\bibliographystyle{naturemag}
\bibliography{ref.bib}

\end{document}